\documentclass[prd,aps,showpacs,nofootinbib,preprintnumbers,onecolumn]{revtex4-1}
\usepackage{graphicx,graphics,epsfig,epstopdf,subfigure,color,psfrag}
\usepackage{hyperref}
\newcommand{\be}{\begin{equation}}
\newcommand{\ee}{\end{equation}}
\newcommand{\bea}{\begin{eqnarray}}
\newcommand{\eea}{\end{eqnarray}}
\newcommand{\nn}{\nonumber}

\begin{document}

\preprint{BARI-TH/670-2013}
\title{ On the anomalous enhancement observed in $B \to D^{(*)} \, \tau \, {\bar \nu}_\tau $ decays}
\author{P. Biancofiore$^{a,b}$, P. Colangelo$^b$ and  F. De Fazio$^b$}
\affiliation{
$^a$Dipartimento di Fisica,  Universit$\grave{a}$  di Bari, Italy\\
$^b$Istituto Nazionale di Fisica Nucleare, Sezione di Bari, Italy}

\begin{abstract}
The BaBar measurements of the ratios ${\cal R}(D^{(*)})=\frac{{\cal B}(B \to D^{(*)} \tau {\bar \nu}_\tau)}{{\cal B}(B \to D^{(*)} \mu {\bar \nu}_\mu)}$  deviate from the standard model expectation, while new results
on the purely leptonic $B \to  \tau {\bar \nu}_\tau$ mode show a better consistency with the standard model, within the uncertainties.  In a  new physics scenario, one possibility to accomodate these two experimental facts consists in considering an
additional tensor operator in the effective weak hamiltonian. We study the effects of such an operator in a set of observables, in semileptonic $B \to D^{(*)}$ modes as well as in  semileptonic $B$ and $B_s$ decays to excited positive  parity charmed  mesons.
\end{abstract}

\pacs{}
\maketitle

\section{Introduction}

The BaBar  measurements of the rates of   $B^-$ and ${\bar B}^0$ semileptonic decays into $D^{(*)}$ and a $\tau$ lepton seem to indicate a significant deviation from the standard model (SM) expectation.
The experimental results concern the  $B\to D^{(*)} \tau {\bar \nu}_\tau $ decay widths normalized  to the widths of  the corresponding modes having a light  $\ell=e,\,\mu$  lepton in the final state \cite{Lees:2012xj}:
\bea
{\cal R}^-(D)=\frac{{\cal B}(B^- \to D^0 \tau^- \,{\bar \nu}_\tau)}{{\cal B}(B^- \to D^0 \ell^- \,{\bar \nu}_\ell)}=0.429 \pm 0.082 \pm 0.052  \,\,\, , &\hskip 0.1 cm &
{\cal R}^-(D^*)=\frac{{\cal B}(B^- \to D^{*0} \tau^- \,{\bar \nu}_\tau)}{{\cal B}(B^- \to D^{*0} \ell^- \, {\bar \nu}_\ell)}=0.322 \pm 0.032 \pm 0.022\,\, , \nn \\
{\cal R}^0(D)=\frac{{\cal B}({\bar B}^0 \to D^+ \tau^- \, {\bar \nu}_\tau)}{{\cal B}({\bar B}^0 \to D^+ \ell^- \, {\bar \nu}_\ell)}=0.469 \pm 0.084 \pm 0.053 \,\,\, , & \hskip 0.1 cm &
{\cal R}^0(D^*)=\frac{{\cal B}({\bar B}^0 \to D^{*+} \tau^- \, {\bar \nu}_\tau)}{{\cal B}({\bar B}^0 \to D^{*+} \ell^- \, {\bar \nu}_\ell)}=0.355 \pm 0.039 \pm 0.021 \,\,\nn \\ \label{data}
\eea
(the first and second error are the statistic and systematic uncertainty, respectively). The measurements  have been estimated to deviate at the global  level of 3.4$\sigma$ with respect to  SM predictions   \cite{Lees:2012xj,Fajfer:2012vx}.
 Therefore, there is the possibility  that  semileptonic processes involving heavy quarks and the $\tau$ lepton are unveiling the effects of particles with large couplings to the heavier fermions, as it is natural for  charged scalars  which could contribute to the tree-level $b \to c \ell \bar \nu$ transitions
\cite{Fajfer:2012vx,Fajfer:2012jt,Becirevic:2012jf,Datta:2012qk,Celis:2012dk,Crivellin:2012ye,Choudhury:2012hn,Tanaka:2012nw}.

\begin{figure}[!h]
 \centering
\includegraphics[width = 0.35\textwidth]{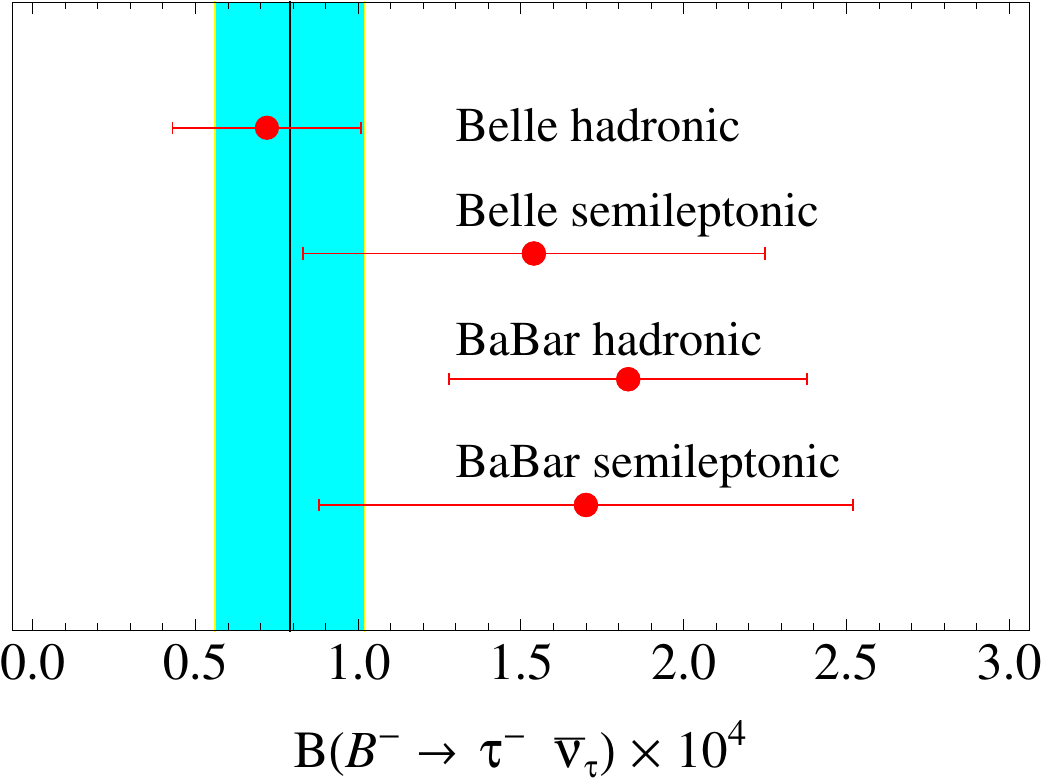}
\caption{Experimental results for ${\cal B}(B^- \to \tau^- {\bar \nu}_\tau)$ \cite{Aubert:2009wt,Hara:2010dk,Adachi:2012mm,Lees:2012ju} together with the SM expectation corresponding to $|V_{ub}|=0.0035 \pm 0.0005$ (vertical band).}\label{fig:Bleptonic}
\end{figure}

Before the observation of  these possible hints of  new physics (NP) in semileptonic  $b \to c$ decays,  the first experimental analyses of   the purely leptonic  $B^- \to \tau^- {\bar \nu}_\tau$ mode also reported an excess of events. In SM
 the   ${\cal B}(B^- \to \tau^- {\bar \nu}_\tau)$  branching fraction is given  by
\be
{\cal B}(B^- \to \tau^- {\bar \nu}_\tau)={G_F^2 m_B m_\tau^2  \over 8 \pi}  \left(1- {m_\tau^2 \over m_B^2}\right)^2f_B^2 \, |V_{ub}|^2\, \tau_{B^-}   \,\,, \label{brBtaunutau}
\ee
neglecting a tiny electromagnetic radiative correction.
Using the lattice QCD average for the $B$ decay constant $f_B=(190.6 \pm 4.7)$ MeV quoted in \cite{Laiho:2009eu}, and varying the  Cabibbo-Kobayashi-Maskawa (CKM) matrix element $|V_{ub}|$ in the range determined from inclusive and exclusive $B$ decays: $|V_{ub}|=0.0035 \pm 0.0005$, the prediction follows: ${\cal B}(B^- \to \tau^- {\bar \nu}_\tau)=(0.79 \pm 0.23)\times 10^{-4}$, in agreement with the outcome of   CKM  matrix fits \cite{Bona:2009cj,Lenz:2010gu}.  This value is  smaller by about a factor of 2  than the experimental results
reported in \cite{Ikado:2006un,Hara:2010dk,Aubert:2007xj,Aubert:2009wt} and compiled in \cite{Rosner:2012np}: ${\cal B}(B^- \to \tau^- {\bar \nu}_\tau)=(1.68 \pm 0.31)\times 10^{-4}$. However,  new  Belle \cite{Adachi:2012mm} and BaBar \cite{Lees:2012ju} measurements,  obtained  using the hadronic tagging method,
\bea
{\cal B}(B^- \to \tau^- {\bar \nu}_\tau)&=&  \left(0.72^{+0.27}_{-0.25}\pm 0.11\right) \times 10^{-4} \,\,\,\,\,  {\rm (Belle)} \nn \\
{\cal B}(B^- \to \tau^- {\bar \nu}_\tau)&=&  \left(1.83^{+0.53}_{-0.49} \pm 0.24 \right) \times 10^{-4} \,\,\,\,\,  {\rm (BaBar)}
\eea
 are more consistent with SM, and  draw the  average  ${\cal B}(B^- \to \tau^- {\bar \nu}_\tau)$ to a smaller value: ${\cal B}(B^- \to \tau^- {\bar \nu}_\tau)=(1.12 \pm 0.22)\times 10^{-4}$,  after the combination with the  semileptonic tagging method results, see fig.\ref{fig:Bleptonic}.

The different trend of the measurements involving $\tau$  in $B$ leptonic and semileptonic decay modes poses two questions. The first one concerns the level of accuracy of the SM predictions for the ratios in (\ref{data}). The second one is which kind of new physics effects, if any, could modify the ratios (\ref{data}) without  affecting the purely leptonic mode.  Indeed,
several  analyses  devoted to try to explain the  anomalies in $B\to D^{(*)} \tau {\bar \nu}_\tau $ within new physics scenarios have considered as possible candidates  models with new scalars  having couplings to leptons proportional to the lepton mass, to guarantee the enhancement of the $\tau$ modes. This is the case of models with two Higgs doublets (2HDM), the best known example being the minimal supersymmetric standard model in which two Higgs doublets are required to give mass to down-type quarks and charged leptons in one case, and up-type quarks in the other. In this framework,  the ratios (\ref{data}) depend on   the mass of the charged Higgs  $H^\pm$ and the ratio $\beta$ of the two Higgs doublet VEVs, and no choice of  such parameters allows to simultaneously reproduce the experimental data on ${\cal R}(D)$ and ${\cal R}(D^*)$  \cite{Lees:2012xj}. Variants of the 2HDM  \cite{Crivellin:2012ye,Celis:2012dk}, together  with other models providing explicit flavour violation \cite{Fajfer:2012jt}, might explain the measurements (\ref{data}); however,  an enhancement of the purely leptonic $B$ decay rate is generally  implied.

In this paper we reconsider both the above mentioned issues. We reanalyse the SM prediction for  $B \to D^{(*)} \ell {\bar \nu}_\ell$, specifying the main sources of uncertainties and  possible improvements. Our results confirm that the most significant deviation is for   ${\cal R}(D^*)$.
Then,  we scrutinize  the effects of  possible NP contributions in the effective weak Hamiltonian having a structure able to affect the ratios (\ref{data}) but leaving the pure $B$ leptonic modes unchanged.  In particular, we focus on a NP operator
constructed from tensor quark and lepton currents. Such a kind of operators have  been also investigated in \cite{Becirevic:2012jf} and \cite{Tanaka:2012nw}, but we devote the main attention to  differential distributions, namely the lepton forward-backward differential asymmetries, in which the sensitivity to the new Dirac structure is maximal, as emphasized in \cite{Datta:2012qk} for different operators. Although there are scenarios in which tensor operators are generated,
in our analysis  we do not rely on explicit models:  our purpose is to identify   physical observables having a mild sensitivity to  hadronic uncertainties,  which therefore can be used to unveil effects   easier to interpret.
It is only worth  mentioning  that these operators emerge,
for example, in models with new coloured bosons  carrying both lepton and baryon quantum number (referred to as leptoquarks, LQ):  SU(5)$_{GUT}$ \cite{Georgi:1974sy}, Pati-Salam SU(4) \cite{Pati:1974yy}, composite  \cite{Schrempp:1984nj}, superstrings \cite{Hewett:1988xc} and technicolor models \cite{Dimopoulos:1979es}.  In the most general formulation of these
models scalar operators may also occur. Leptoquarks  couple to  quarks and leptons and, from limits on flavour changing neutral currents,  preferably to those within the same SM generation. Searches for leptoquarks decaying to 2$\tau$ and 2$b$ jets, performed by the CMS Collaboration at the CERN LHC,  bound (preliminarly)   the mass of a possible scalar leptoquark to $M(LQ)>525$ GeV,  and to  $M(LQ)>760$  GeV for a vector leptoquark \cite{cms};   other  bounds can be found in \cite{leptoquarks}.

 In our analysis of semileptonic $B$ decays, we first consider $D$ and $D^*$ mesons in the final state, and then turn to  the interesting case of  final states with excited positive parity charmed mesons.

\section{Exclusive $b \to c \ell {\bar \nu}_\ell$ Decays}
 We consider the $b \to c \ell {\bar \nu}_\ell$ effective hamiltonian   comprising  the  SM term and an additional operator     \cite{Becirevic:2012jf,Tanaka:2012nw}:
\be
H_{eff}=H_{eff}^{SM}+H_{eff}^{NP} =  {G_F \over \sqrt{2}}V_{cb} \left[ {\bar c} \gamma_\mu (1-\gamma_5) b \, {\bar \ell} \gamma^\mu (1-\gamma_5) {\bar \nu}_\ell + \epsilon_T^\ell \, {\bar c} \sigma_{\mu \nu} (1-\gamma_5) b \, {\bar \ell} \sigma^{\mu \nu} (1-\gamma_5) {\bar \nu}_\ell \right] \,\,\, . \label{heff}
\ee
$G_F$ is the Fermi constant and  $V_{cb}$  the CKM matrix element.  $\epsilon_T^\ell$ is the relative complex coupling of the new tensor term with respect to the SM one. It is assumed that the main coupling is to the heaviest lepton, hence we set $\epsilon_T^\ell=0$ for $\ell=e,\mu$  and  $\epsilon_T\equiv \epsilon_T^\tau$. This coupling can be bound experimentally, so that the effects of the new operator can be scrutinized
in physical observables which, in general,  are expressed  as a SM, a new physics and an interference contribution.
For example, the differential  $B(p) \to M_c(p^\prime) \ell(p_1) {\bar \nu}_\ell(p_2)$ decay rate, with  $M_c$  a charmed meson,   reads:
\be
{d \Gamma \over dq^2}(B \to M_c \ell \bar \nu_\ell)=C(q^2) \left[ {d \tilde \Gamma \over dq^2}(B \to M_c \ell \bar \nu_\ell)\Big|_{SM}+{d \tilde \Gamma \over dq^2}(B \to M_c \ell \bar \nu_\ell)\Big|_{NP}+{d \tilde \Gamma \over dq^2}(B \to M_c \ell \bar \nu_\ell)\Big|_{INT} \right]\,, \label{dgammadq2-generic}
\ee
with $q=p-p^\prime$ and $C(q^2)$ defined as
\be
C(q^2)={G_F^2 |V_{cb}|^2 \lambda^{1/2}(m_B^2,m_{M_c}^2,q^2) \over 192 \pi^3 m_B^3}
\left(1 -{m_\ell^2 \over q^2 } \right)^2 \,\,\, ; \label{C-factor}
\ee
$\lambda(x,y,z)=x^2+y^2+z^2-2(xy+xz+yz)$ is the triangular function.
To compute the three terms in (\ref{dgammadq2-generic})  we need the relevant  hadronic matrix elements.

\subsection{$B \to D \ell {\bar \nu}_\ell$ }
 The  hadronic  matrix elements in $B \to D \ell {\bar \nu}_\ell$ can be parametrized in a standard way,
\bea
<D(p^\prime)|{\bar c} \gamma_\mu b| B(p)>&=&F_1(q^2)(p+p^\prime)_\mu+{m_B^2-m_D^2 \over q^2} \left[F_0(q^2)-F_1(q^2)\right] q_\mu \,\,\, , \label{V-A-D} \\
<D(p^\prime)|{\bar c} \sigma_{\mu \nu}(1-\gamma_5) b| B(p)>&=&{F_T(q^2) \over m_B+m_D} \, \epsilon_{\mu \nu \alpha \beta} p^{\prime \alpha} p^\beta+i\,{G_T(q^2) \over m_B+m_D} \, (p_\mu p^\prime_\nu-p_\nu p^\prime_\mu)\,, \label{T-D}
\eea
(with $F_T=G_T$  from the relation $\sigma_{\mu \nu} \gamma_5=\frac{i}{2}\epsilon_{\mu \nu \alpha \beta}\,\sigma^{\alpha \beta}$),
so that  the three terms in (\ref{dgammadq2-generic}) read:
\bea
{d \tilde \Gamma \over dq^2}(B \to D \ell \bar \nu_\ell)\Big|_{SM} &=&
\lambda (m_B^2,m_{D}^2,q^2)\left(1+{m_\ell^2 \over 2 q^2} \right) \left[F_1(q^2)\right]^2+m_B^4 \left(1-{m_D^2 \over  m_B^2} \right)^2{3  m_\ell^2 \over 2 q^2} \left[F_0(q^2)\right]^2 \,, \label{D-SM}
\\
{d \tilde \Gamma \over dq^2}(B \to D \ell \bar \nu_\ell)\Big|_{NP} &=&{|\epsilon_T|^2 \over 2} {q^2 \over (m_B+m_D)^2}\, \lambda (m_B^2,m_{D}^2,q^2) \left(1+2{m_\ell^2 \over q^2} \right) \, \left[F_T(q^2)+G_T(q^2)\right]^2  \label{D-NP} \,, \\
{d \tilde \Gamma \over dq^2}(B \to D \ell \bar \nu_\ell)\Big|_{INT}& =&-3 Re[\epsilon_T]{ m_\ell \over m_B + m_D}\, \lambda (m_B^2,m_{D}^2,q^2) \, F_1(q^2)\, \left[F_T(q^2)+G_T(q^2)\right] \label{D-INT}\,.
\eea
In the infinite heavy quark mass limit, formalized by the heavy quark effective theory (HQET),     the form factors in (\ref{V-A-D}-\ref{T-D}) can all be related to the Isgur-Wise function $\xi$ \cite{Isgur:1989ed}. The result is   known \cite{hqet,hqet1}:
expressing  $F_1(q^2)$ and $F_0(q^2)$ in terms of  two other form factors $h_+(w)$ and $h_-(w)$:
\bea
F_1(q^2)&=& \frac{1}{ 2 \sqrt{m_B m_D}} \left[(m_B+m_D) h_+(w) - (m_B-m_D)h_-(w)\right] \\
\frac{m_B^2 -m_D^2}{q^2}\left[F_0(q^2)-F_1(q^2)\right]&=& \frac{1}{ 2 \sqrt{m_B m_D}} \left[(m_B+m_D) h_-(w) - (m_B-m_D)h_+(w) \right]\,\,,
\eea
and defining the meson momenta in terms of four-velocities, $p=m_B v$ and $p^\prime=m_D v^\prime$, with $w=v \cdot v^\prime$ and  $q^2 = m_B^2+m_D^2-2m_B m_D w$,
at the leading order in the heavy quark   and  $\alpha_s$ expansion one has
\be
h_+(w)=\xi(w) \,\,\,\, , \hskip 1 cm h_-(w)=0 \,\,\, , \label{hpiu-meno-xi}
\ee
with $\xi(w)$ the Isgur-Wise function.
Also the form factors in  (\ref{T-D}) are related to  $\xi(w)$ at the same order expansion:
\be
F_T(q^2)=G_T(q^2)={m_B+m_D \over \sqrt{m_B m_{D}}} \,\xi(w) \,\,\, .  \label{FT-GT-xi}
\ee

\begin{figure}[!t]
\centering
\includegraphics[width = 0.4\textwidth]{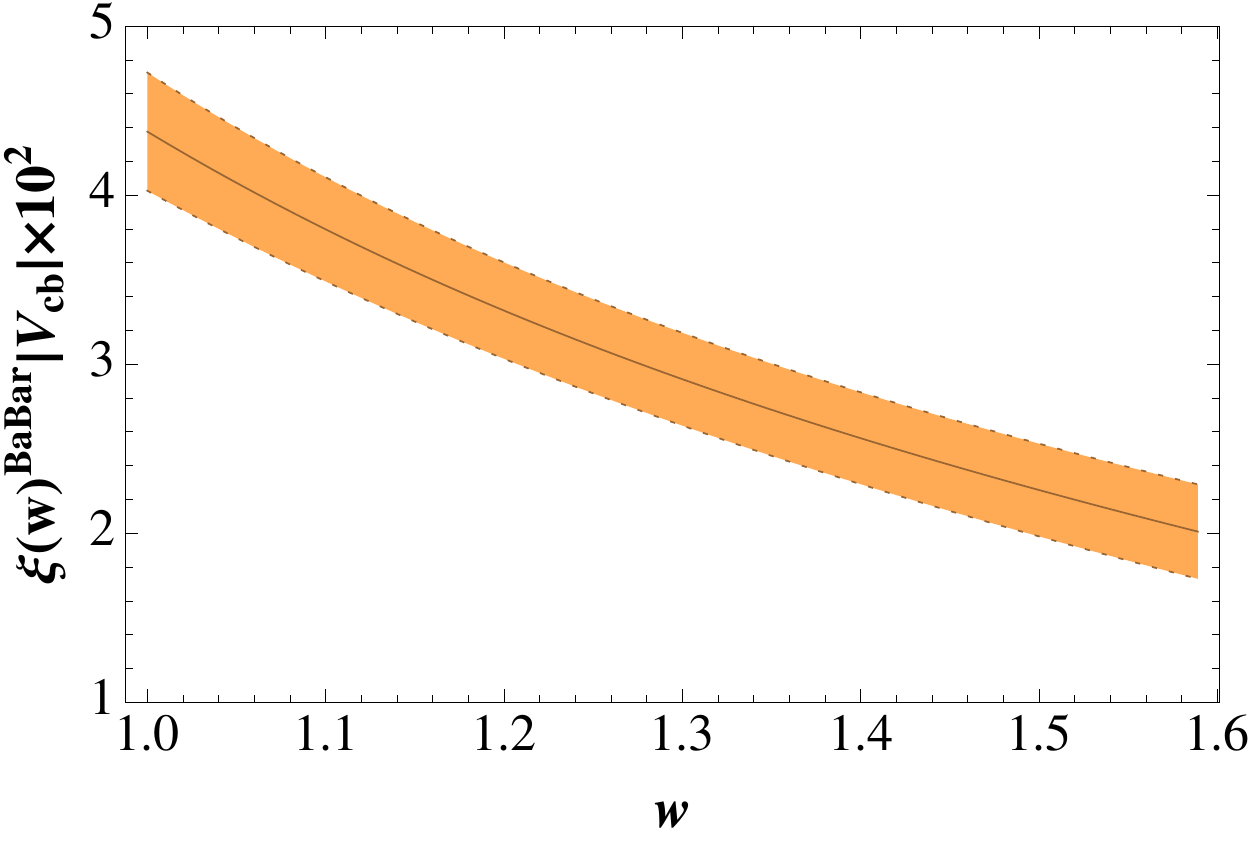}\hspace*{0.5cm}
\includegraphics[width = 0.4\textwidth]{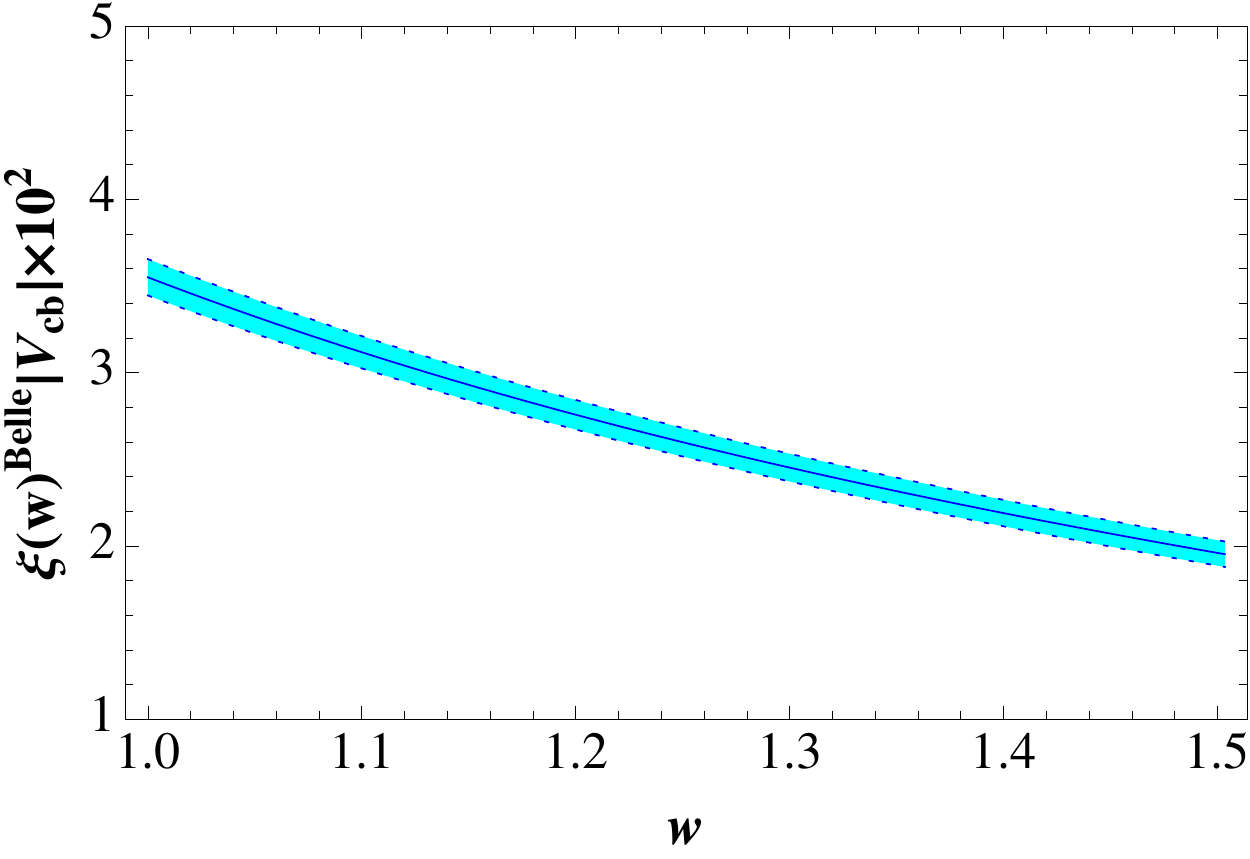}
\caption{Isgur-Wise function $\xi(w)$ (times $|V_{cb}|\times 10^2$) obtained using the  BaBar  data on  ${\bar B}^0 \to D^+ \ell^- \bar \nu_\ell$ (left) and  the Belle data on ${\bar B}^0 \to D^{*+} \ell^- \bar \nu_\ell$ (right) . The width of the curves is due to  the errors in the parameters fitted in the two  cases  and to the uncertainty on  ${\bar \Lambda}$ and $\alpha_s$ in the determination of the form factor.}\label{fig:xi}
\end{figure}

At the next-to-leading order, corrections must be taken into account, which at first are needed  for the
study  of  the  decay  in   SM. We elaborate a determination of the functions $h_+$, $h_-$ and $\xi $ based on a combination of experimental and theoretical information. The experimental input comes from the
 BaBar  analysis of  $B \to D \mu {\bar \nu}_\mu$ \cite{Aubert:2009ac},  the differential  rate of which, neglecting the lepton mass,  reads:
\be
{d \Gamma \over dw}( B \to D \ell \bar \nu_\ell)={G_F^2 |V_{cb}|^2 \over 48 \pi^3 }  m_B^5 r^3(1+r)^2 (w^2-1)^{3/2} [F_D(w)]^2 \,\,\, ,
\ee
with
\be
F_D(w)= \left[ h_+(w)-{1-r \over 1+r} h_-(w) \right] \ee
and $\displaystyle r=\frac{m_D}{m_B}$.  Using the parametrization \cite{Caprini:1997mu}
\be
F_D(w)=F_D(1) \Big\{ 1-8 \rho_1^2 z+(51 \rho_1^2-10)z^2-(252\rho_1^2-84)z^3 \Big\} \,\,\,  \label{FD}
\ee
in terms of the variable
\be
z={ \sqrt{w+1}-\sqrt{2} \over \sqrt{w+1}+\sqrt{2}} \,\,\, , \label{zeta}
\ee
from the fit of the product $G^{BaBar}(w)=F_D(w)|V_{cb}|$ the BaBar Collaboration provides  the  parameters $G^{BaBar}(1)=F_D(1)|V_{cb}|$ and $\rho_1^2$.
The outcome of the fit is slightly different for  $B^-$ or ${\bar B}^0$ modes;  we consider for definiteness the  ${\bar B}^0$ case  \cite{Aubert:2009ac}
\footnote{The   average between the charged and neutral $B$ decay modes is quoted as   $G^{BaBar}(1)=(42.3 \pm 1.9 \pm 1.4) \, 10^{-3}$ , $ \rho_1^2=1.20 \pm 0.09 \pm 0.04$.},
\be
G^{BaBar}(1)=(44.9 \pm 3.2 \pm 1.6)  \, 10^{-3}\,\,\, , \hskip 1cm \rho_1^2=1.29 \pm 0.14 \pm 0.05\,\,.
\ee
This result can be  translated into  a determination of $\xi(w)$,   expressing  the form factors $h_\pm(w)$ in terms of the Isgur-Wise function  and including the   $\alpha_s$ and $1/m_{b,c}$ corrections  worked out by M. Neubert in \cite{hqet} and  by I. Caprini et al., in \cite{Caprini:1997mu}:
\bea
h_+(w)&=&\left[C_1 +{w+1 \over 2} (C_2+C_3) +(\epsilon_b+\epsilon_c) L_1 \right] \xi(w)={\tilde h}_+(w) \, \xi(w) \label{hpiu} \\
h_-(w)&=&\left[{w+1 \over 2} (C_2-C_3) +(\epsilon_c -\epsilon_b) L_4 \right] \xi(w)={\tilde h}_-(w) \, \xi(w) \label{hmeno}
\eea
with $\epsilon_b=\displaystyle{1 \over 2 m_b}$, $\epsilon_c=\displaystyle{1 \over 2 m_c}$.
The  coefficients $C_{1,2,3}$ and   $L_i$ are collected in appendix \ref{app:coefficients}. $C_i$ account for the perturbative corrections,   $L_i$ for  the heavy quark mass corrections and depend on the hadronic parameter   $\bar \Lambda$, the difference between the heavy meson ($B,\,D$) and the heavy quark ($b,\,c$) mass in the heavy quark limit.
We use  $m_b=4.8$ GeV and $m_c=1.4$  GeV and   a conservative  value  ${\bar \Lambda}=0.5 \pm 0.2$ GeV \cite{hqet},  so that  the uncertainty  in  ${\bar \Lambda}$ encompasses the error on
${\bar \Lambda}/m_b$ and ${\bar \Lambda}/m_c$.
The  Isgur-Wise function $\xi(w)$  resulting from
\be
|V_{cb}| \,  \xi(w)=\frac{G^{BaBar}(w)}{\left[ \tilde h_+(w)-{1-r \over 1+r} \, \tilde h_-(w) \right]} \,\,
\label{xi}
\ee
 is depicted in  fig.\ref{fig:xi} (left panel).

The form factors  needed for analysis of the mode with  $\tau$  can be separately derived using again  Eqs.(\ref{hpiu},\ref{hmeno}):
\bea
|V_{cb}| \,  h_+(w)&=& \frac{1}{1- {1-r \over 1+r} \, A(w)} \,\, G^{BaBar}(w)  \\
|V_{cb}| \,  h_-(w)&=& \frac{A(w)}{1- {1-r \over 1+r} \, A(w)} \,\, G^{BaBar}(w)
\eea
with $A=\tilde h_-/\tilde h_+$.  For the matrix elements of the tensor operator,  we  use $\xi(w)$ also in (\ref{FT-GT-xi}). In the standard model, the  results for  the semileptonic
${\bar B}^0 \to D^+$ branching fractions can  be quoted as
\bea
{\cal B}({\bar B}^0 \to D^+ \ell^- {\bar \nu}_\ell)\Big|_{SM} &=& (2.15 \pm 0.45) \times 10^{-2} \\
{\cal B}({\bar B}^0 \to D^+ \tau^- {\bar \nu}_\tau)\Big|_{SM} &=& (0.70 \pm 0.12) \times 10^{-2}
 \eea
and,  taking the correlation between the   predictions  for $\ell$ and $\tau$ into account,
 \be
 {\cal R}^0(D)\Big|_{SM}=\frac{{\cal B}({\bar B}^0 \to D^+ \tau^- {\bar \nu}_\tau)}{{\cal B}({\bar B}^0 \to D^+ \ell^- {\bar \nu}_\ell) }\Big|_{SM}=0.324 \pm 0.022 \,\,\, . \label{R0res}
 \ee
The SM prediction for ${\cal R}^0(D)$ deviates from  the  measurement  (\ref{data}) (with  statistic and systematic uncertainties combined in quadrature) by about  1.5$\sigma$ .
The deviation   is smaller  in the charged  ${\cal R}^-(D)$ case.

The stability of (\ref{R0res}) against changes of the input information on form factors is noticeable: sensitivity to  $1/m_Q$ corrections can be estimated varying $\bar \Lambda$, and this modifies the central value
at a few per mille level. Sensitivity to the radiative corrections can be assessed  changing the  scale in $\alpha_s$   as indicated in appendix \ref{app:coefficients}, and  also these corrections are not effective.  Since the value at zero recoil $G^{BaBar}(1)$ cancels out in the ratio,
the main uncertainty  in  (\ref{R0res}) comes from  the error on the parameter $\rho_1^2$ experimentally determined. The value of ${\cal R}^0(D)$ coincides with the one obtained  using the form factors $F_1$ and $F_0$ from  lattice QCD with finite quark masses \cite{Becirevic:2012jf}.

\subsection{$B \to D^* \ell {\bar \nu}_\ell$}
While the results  for ${\cal R}^0(D)$ and ${\cal R}^-(D)$ do not display a statistically significant deviation from the SM expectation,  the case of ${\cal R}^0(D^*)$,  ${\cal R}^-(D^*)$ is  quite different.
The standard parameterization of the   $B \to D^*$   matrix element  in terms of form factors is
\bea
<D^*(p^\prime,\epsilon)|{\bar c} \gamma_\mu(1-\gamma_5) b| {\bar B}(p)>&=&- {2 V(q^2) \over m_B+m_{D^*}} i \epsilon_{\mu \nu \alpha \beta} \epsilon^{*\nu}  p^\alpha p^{\prime \beta} -\Bigg\{ (m_B+m_{D^*}) \left[ \epsilon^*_\mu -{(\epsilon^* \cdot q) \over q^2} q_\mu \right] A_1(q^2) \nn\\
&&- {(\epsilon^* \cdot q) \over  m_B+m_{D^*}} \left[ (p+p^\prime)_\mu -{m_B^2-m_{D^*}^2 \over q^2} q_\mu \right] A_2(q^2)
+ (\epsilon^* \cdot q){2 m_{D^*} \over q^2} q_\mu A_0(q^2) \Bigg\} \nn \\ \label{FF-D*-mio}
\eea
(with the condition  $\displaystyle A_0(0)= \frac{m_B + m_{D^*}}{2 m_{D^*}} A_1(0) -  \frac{m_B - m_{D^*}}{2 m_{D^*}}  A_2(0)$) and
\bea
<D^*(p^\prime,\epsilon)|{\bar c} \sigma_{\mu \nu}(1-\gamma_5) b| {\bar B}(p)>&=&T_0(q^2) {\epsilon^* \cdot q \over (m_B+ m_{D^*})^2} \epsilon_{\mu \nu \alpha \beta} p^\alpha p^{\prime \beta}+
T_1(q^2) \epsilon_{\mu \nu \alpha \beta} p^\alpha \epsilon^{*\beta}+T_2(q^2) \epsilon_{\mu \nu \alpha \beta} p^{\prime \alpha} \epsilon^{*\beta}\nn \\
&+&i \, \Big[ T_3(q^2) (\epsilon^*_\mu p_\nu -\epsilon^*_\nu p_\mu)+T_4(q^2) (\epsilon^*_\mu p^\prime_\nu -\epsilon^*_\nu p^\prime_\mu) \nn \\
&+&T_5(q^2) {\epsilon^* \cdot q \over (m_B+ m_{D^*})^2}(p_\mu p^\prime_\nu -p_\nu p^\prime_\mu)\Big]  \,\,,  \label{mat-tensor-Dstar}
\eea
with $\epsilon$  the $D^*$ polarization vector.
We choose the  helicity basis for  $D^*$
\be
\epsilon^\mu_L = {1 \over m_{D^*}} \left( |\vec p^\prime|, 0,0, E^\prime \right) \,\,\,\, , \hspace*{0.5cm}
\epsilon^\mu_\pm={1 \over \sqrt{2}} \left(0,1, \mp i, 0 \right)\,\,\,\, ,  \label{pol-vec}
\ee
with  $E^\prime$ and   $\vec p^\prime$
the $D^*$ energy  and three-momentum in the $B$ rest frame  ($E^\prime=\sqrt{m_{D^*}^2+|\vec p^\prime|^2}$ and $\displaystyle |\vec p^\prime|=\lambda (m_B^2,m_{D^*}^2,q^2)/2 m_B$).
 The conditions $\epsilon^\mu_a \cdot p^\prime=0$ and $\epsilon^\mu_a \cdot \epsilon_{\mu,b}=-\delta_{ab}$, with $a,b=L,\pm$, are fulfilled.
The  differential decay rates for  the   longitudinal and  the  transverse $D^*$  polarization   in terms of form factors are obtained from
\bea
{d \tilde \Gamma_L \over dq^2}(B \to D^* \ell \bar \nu_\ell)\Big|_{SM} &=&
\frac{1}{4m_{D^*}^2}\Bigg\{6\lambda(m_B^2,m_{D^*}^2,q^2) m_{D^*}^2 \frac{m_\ell^2}{q^2}[A_0(q^2)]^2 \nn \\
&&+\left(1+\frac{m_\ell^2}{2q^2} \right) \left[(m_B+m_{D^*})(m_B^2-m_{D^*}^2-q^2)A_1(q^2)-\frac{\lambda(m_B^2,m_{D^*}^2,q^2)}{m_B+m_{D^*}}A_2(q^2) \right]^2 \Bigg\} \,\, ,
 \label{DstarSML} \\
{d \tilde \Gamma_L \over dq^2}(B \to D^* \ell \bar \nu_\ell)\Big|_{NP} & = &|\epsilon_T|^2 \frac{q^2}{8}\left(1+\frac{2m_\ell^2}{q^2} \right) \Big[\frac{\lambda(m_B^2,m_{D^*}^2,q^2)}{m_{D^*}(m_B+m_{D^*})^2}{\tilde T}_0(q^2)+2\frac{m_B^2+m_{D^*}^2-q^2}{m_{D^*}}{\tilde T}_1(q^2)+4m_{D^*}{\tilde T}_2(q^2) \Big]^2, \nn \\
\label{DstarNPL} \\
{d \tilde \Gamma_L \over dq^2}(B \to D^* \ell \bar \nu_\ell)\Big|_{INT} &=&-Re(\epsilon_T)\frac{3m_\ell}{4(m_B+m_{D^*})}\Big[(m_B+m_{D^*})^2(m_B^2-m_{D^*}^2-q^2)A_1(q^2)-\lambda(m_B^2,m_{D^*}^2,q^2)A_2(q^2) \Big] \nn\\
&&\left[\frac{\lambda(m_B^2,m_{D^*}^2,q^2)}{m_{D^*}^2(m_B+m_{D^*})^2}{\tilde T}_0(q^2)+\frac{2(m_B^2+m_{D^*}^2-q^2)}{m_{D^*}^2}{\tilde T}_1(q^2)+4{\tilde T}_2(q^2) \right] \,\, , \label{DstarINTL}
\eea
\bea
{d \tilde \Gamma_\pm \over dq^2}(B \to D^* \ell \bar \nu_\ell)\Big|_{SM} &=& q^2\left(1+\frac{m_\ell^2}{2q^2} \right) \Bigg\{(m_B+m_{D^*})^2[A_1(q^2)]^2+\frac{\lambda(m_B^2,m_{D^*}^2,q^2)}{(m_B+m_{D^*})^2}[V(q^2)]^2\Bigg\}  \,\, ,\label{DstarSMpm} \\
{d \tilde \Gamma_\pm \over dq^2}(B \to D^* \ell \bar \nu_\ell)\Big|_{NP} &=&|\epsilon_T|^2 \left(1+\frac{2m_\ell^2}{q^2} \right)
\Bigg\{\lambda(m_B^2,m_{D^*}^2,q^2)[{\tilde T}_1(q^2)+{\tilde T}_2(q^2)]^2  \nn \\
&&+2q^2\left[m_B^2[{\tilde T}_1(q^2)]^2+m_{D^*}^2[{\tilde T}_2(q^2)]^2+(m_B^2+m_{D^*}^2-q^2){\tilde T}_1(q^2){\tilde T}_2(q^2)\right]\Bigg\} \,\, ,
\label{DstarNPpm} \\
{d \tilde \Gamma_\pm \over dq^2}(B \to D^* \ell \bar \nu_\ell)\Big|_{INT} &=&-Re(\epsilon_T)3m_\ell\Bigg\{2q^2(m_B+m_{D^*})A_1(q^2){\tilde T}_1(q^2)
\nn \\
&&+\left[(m_B+m_{D^*})(m_B^2-m_{D^*}^2-q^2)A_1(q^2)-
\frac{\lambda(m_B^2,m_{D^*}^2,q^2)}{(m_B+m_{D^*})}V(q^2) \right][{\tilde T}_1(q^2)+{\tilde T}_2(q^2)]\Bigg\} \,\,,\nn \\ \label{DstarINTpm}
\eea
 to be multiplied by the factor $C(q^2)$ in  (\ref{C-factor}).
We have used the combinations
\bea
{\tilde T}_0(q^2)&=&T_0(q^2)-T_5(q^2) \nn \\
{\tilde T}_1(q^2)&=&T_1(q^2)+T_3(q^2)  \\
{\tilde T}_2(q^2)&=&T_2(q^2)+T_4(q^2) \,\,. \nn
\eea
At the leading order in the heavy quark expansion,  the form factors in (\ref{FF-D*-mio}) and (\ref{mat-tensor-Dstar}) are related to the Isgur-Wise function, while other contributions appear at the next-to-leading order.  Analogously to the decay to $D$, one expresses $V$ and $A_i$  in terms of  form factors $h_V$ and $h_{A_i}$,
 \bea
 V(q^2)&=& {m_B+m_{D^*} \over 2 \sqrt{m_B m_{D^*}}} h_V(w) \,\,\,  \nn \\
 A_1(q^2) &=& \sqrt{m_B m_{D^*}}{w+1 \over m_B+m_{D^*}} h_{A_1}(w) \,\,\,  \nn \\
 A_2(q^2) &=& { m_B+m_{D^*} \over 2 \sqrt{m_B m_{D^*}}} \left[h_{A_3}(w)+{m_{D^*} \over m_B} h_{A_2}(w)\right]  \,\,\,  \nn \\
A_0(q^2) &=& { 1 \over 2 \sqrt{m_B m_{D^*}}} \Big[ m_B (w+1) h_{A_1}(w) -(m_B-m_{D^*} w)h_{A_2}(w)-(m_Bw-m_{D^*})h_{A_3}(w) \Big] \,\,\,
\eea
with $q^2=m_B^2+m_{D^*}^2-2 m_B m_{D^*} w$.
Including $\alpha_s$  and  $\displaystyle{\frac{1}{m_{b}}}$ and  $\displaystyle{\frac{1}{m_{c}}}$ corrections, the relations have been worked out  \cite{hqet,Caprini:1997mu}:
\bea
h_V(w) &=& \left[ C_1 +\epsilon_c (L_2 -L_5) +\epsilon_b (L_1 -L_4) \right] \, \xi(w) \label{hv}\\
h_{A_1}(w) &=& \left[ C_1^5 +\epsilon_c \left(L_2-{w-1 \over w+1}L_5 \right) +\epsilon_b \left(L_1 -{w-1 \over w+1}L_4 \right)  \right] \,  \xi(w)\label{ha1} \\
h_{A_2}(w) &=& \left[ C_2^5 +\epsilon_c (L_3+L_6) \right] \, \xi(w) \label{ha2}\\
h_{A_3}(w) &=& \left[ C_1^5+C_3^5 +\epsilon_c (L_2 -L_3 -L_5+L_6) + \epsilon_b (L_1 -L_4) \right] \, \xi(w) \,\,\, . \label{ha3}
\eea
The expressions of $C_i$, which incorporate the radiative corrections,  and $L_i$ are  collected in  appendix \ref{app:coefficients}:    the $L_i$ terms account for the ${\cal O}(1/m_Q)$ corrections in the heavy quark expansion, and are  determined from
QCD sum rule analyses of the subleading  form factors \cite{hqet}.
On the other hand, the relations of the form factors $T_i$ in (\ref{mat-tensor-Dstar}) to $\xi(w)$ in the heavy quark limit are:
\bea
T_0(q^2)&=&T_5(q^2)=0 \nn \\
T_1(q^2)&=&T_3(q^2)=\sqrt{m_{D^*} \over m_B} \xi(w) \\
T_2(q^2)&=&T_4(q^2)=\sqrt{ m_B\over m_{D^*} } \xi(w) \,\,\,\ ; \nn
\eea
we use these expressions in the analysis of the tensor operator.

Let us focus on  the SM.
Due to the heavy quark spin symmetry a unique form factor describes both $B \to D$ and $B \to D^*$ transitions, so that we could use the  Isgur-Wise function  found in the previous section. To partially take into account the different experimental systematics,  we  choose to use the determination of $\xi$ obtained by Belle Collaboration from the analysis of  ${\bar B}^0 \to D^{*+}  \mu {\bar \nu}_\mu$  \cite{Dungel:2010uk}, for which the
differential decay rate, neglecting the lepton mass,  is
\be
{d \Gamma \over dw}(B \to D^* \ell \bar \nu_\ell)= {G_F^2 |V_{cb}|^2  \over 48 \pi^3} (m_B-m_{D^*})^2 m_{D^*}^3 {\cal G}(w) {\cal F}^2(w) \,\,\, ,
\ee
with
\bea
{\cal G}(w) {\cal F}^2(w) &=& h_{A_1}^2(w) \sqrt{w^2-1} \, (w+1)^2 \nn \\
&& \left\{2\left[{1-2wr^*+r^{*2} \over (1-r^*)^2} \right] \left[1+R_1(w)^2 {w-1 \over w+1} \right]+\left[1+(1-R_2(w)){w-1 \over 1-r^*} \right]^2 \right\} \,\,\, . \label{eq:G}
\eea
In (\ref{eq:G})  $r^*=\displaystyle{m_{D^*} \over m_B}$,  and ${\cal G}$, $R_1$ and $R_2$ are given by
\bea
{\cal G}(w)&=& \sqrt{w^2-1} (w+1)^2 \left[1+4{w \over w+1}{1-2wr^*+r^{*2} \over (1-r^*)^2}\right] \,\,\, , \nn \\
R_1(w)&=& (R^{*})^2 {w+1 \over 2}{V(w) \over A_1(w)}\,\,\, , \label{R1}\\
R_2(w)&=& (R^{*})^2 {w+1 \over 2}{A_2(w) \over A_1(w)} \,\,\, , \nn
\eea
with  $R^*=2 \displaystyle{\sqrt{m_B m_{D^*}} \over m_B+m_{D^*}}$.
The three unwnown functions in (\ref{eq:G},\ref{R1}) have been determined by  Belle
adopting the parametrization \cite{Caprini:1997mu}
\bea
h_{A_1}(w)&=&h_{A_1}(1) [\, 1-8 \rho^2 z+(53 \rho^2 -15)z^2-(231\rho^2-91)z^3] \label{ha1-belle} \\
R_1(w)&=&R_1(1)-0.12 \, (w-1)+0.05 \, (w-1)^2  \label{R1-belle} \\
R_2(w)&=&R_1(1)+0.11 \, (w-1)-0.06 \, (w-1)^2  \label{R2-belle}
\eea
(with $z$ defined in (\ref{zeta})).  The  fit of the parameters in (\ref{ha1-belle}-\ref{R2-belle}) is quoted as
 \cite{Dungel:2010uk}
 \bea
  {\cal F}(1)|V_{cb}|&=&(34.6 \pm 0.2 \pm 1.0) \times 10^{-3} \,\,  \nn \\
  \rho^2 &=& 1.214 \pm 0.034 \pm 0.009  \,\, \nn \\
  R_1(1)&=& 1.401 \pm 0.034 \pm 0.018 \,\,  \label{belle-par} \\
  R_2(1) &=& 0.864 \pm 0.024 \pm 0.008 \,\, .  \nn
  \eea
From these expressions one can reconstruct $\xi(w)$,
  \be
  h_{A_1}(w)={\tilde h}_{A_1}(w) \, \xi(w)
  \ee
  with  ${\tilde h}_{A_1}$ defined through Eq.(\ref{ha1}).
  The fit provides us  with the  determination  depicted in fig.\ref{fig:xi} (right panel). Through Eqs.(\ref{hv},\ref{ha2},\ref{ha3}) the form factors $h_V$, $h_{A_2}$ and $h_{A_3}$ can be reconstructed including the NLO
  $1/m_Q$ and $\alpha_s$ corrections, and  also  ${\cal B}({\bar B}^0 \to D^{*+} \tau^- \bar \nu_\ell)$ can be computed. The results are:
\bea
{\cal B}({\bar B}^0 \to D^{*+} \ell^- \bar \nu_\ell)\Big|_{SM} &=& (4.62 \pm 0.33 )\times 10^{-2}  \nn\\
{\cal B}({\bar B}^0 \to D^{*+} \tau^- \bar \nu_\tau)\Big|_{SM} &=& (1.16 \pm 0.08 )\times 10^{-2} \,\,  \label{brsSM} \eea
and, taking  the correlation between the predictions  for the $\ell$ and $\tau$  mode into account,
\be
 {\cal R}^0(D^*)\Big|_{SM}=\frac{{\cal B}({\bar B}^0 \to D^{*+} \tau^- {\bar \nu}_\tau)}{{\cal B}({\bar B}^0 \to D^{*+} \ell^- {\bar \nu}_\ell) }\Big|_{SM}=0.250 \pm 0.003 \,\,\,  .\label{eq:RD*SM}
 \ee
The result (\ref{eq:RD*SM}) deviates  from the measurement  in (\ref{data}) (with  statistic and systematic errors combined in quadrature) by  2.3$\sigma$.
It coincides with the one in \cite{Fajfer:2012vx,Celis:2012dk,Tanaka:2012nw},
due to  the stability of the  ratio  ${\cal R}^0(D^*)$  against changes of the input parameters:
 varying  the central value of  $\bar \Lambda$ and of the quark masses by 30$\%$  produces less than $1\%$ variation in the result. The radiative corrections,  changing the scale in $\alpha_s$ as
mentioned in appendix \ref{app:coefficients}, do not  produce  an appreciable variation of the result.
On the other hand,  in the individual branching fractions  there is a mild sensitivity  to $\bar \Lambda$: setting this parameter to zero (i.e. ignoring $1/m_Q$ corrections)   the branching fractions in (\ref{brsSM}) are reduced by about $5\%$.  In the charged case,   there is a deviation of  1.8$\sigma$ between  the  SM prediction for ${\cal R}(D^*)$ and  the measurement in (\ref{data}).

\section{Effects of the tensor operator on ${\cal R}(D^{(*)})$ and other observables}
\begin{figure}[!b]
\centering
\includegraphics[width = 0.35\textwidth]{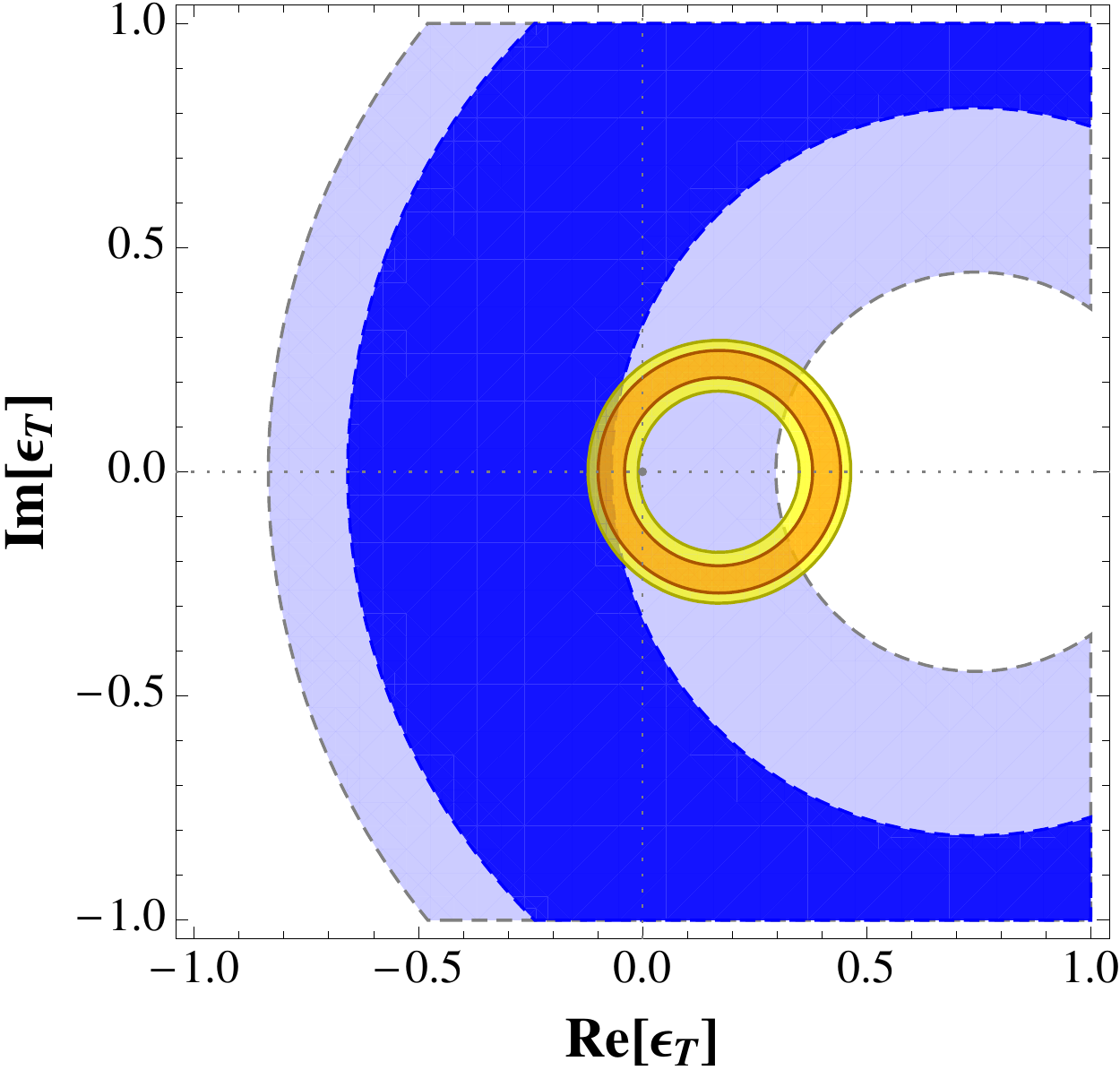} \hspace*{0.5cm}
\includegraphics[width = 0.35\textwidth]{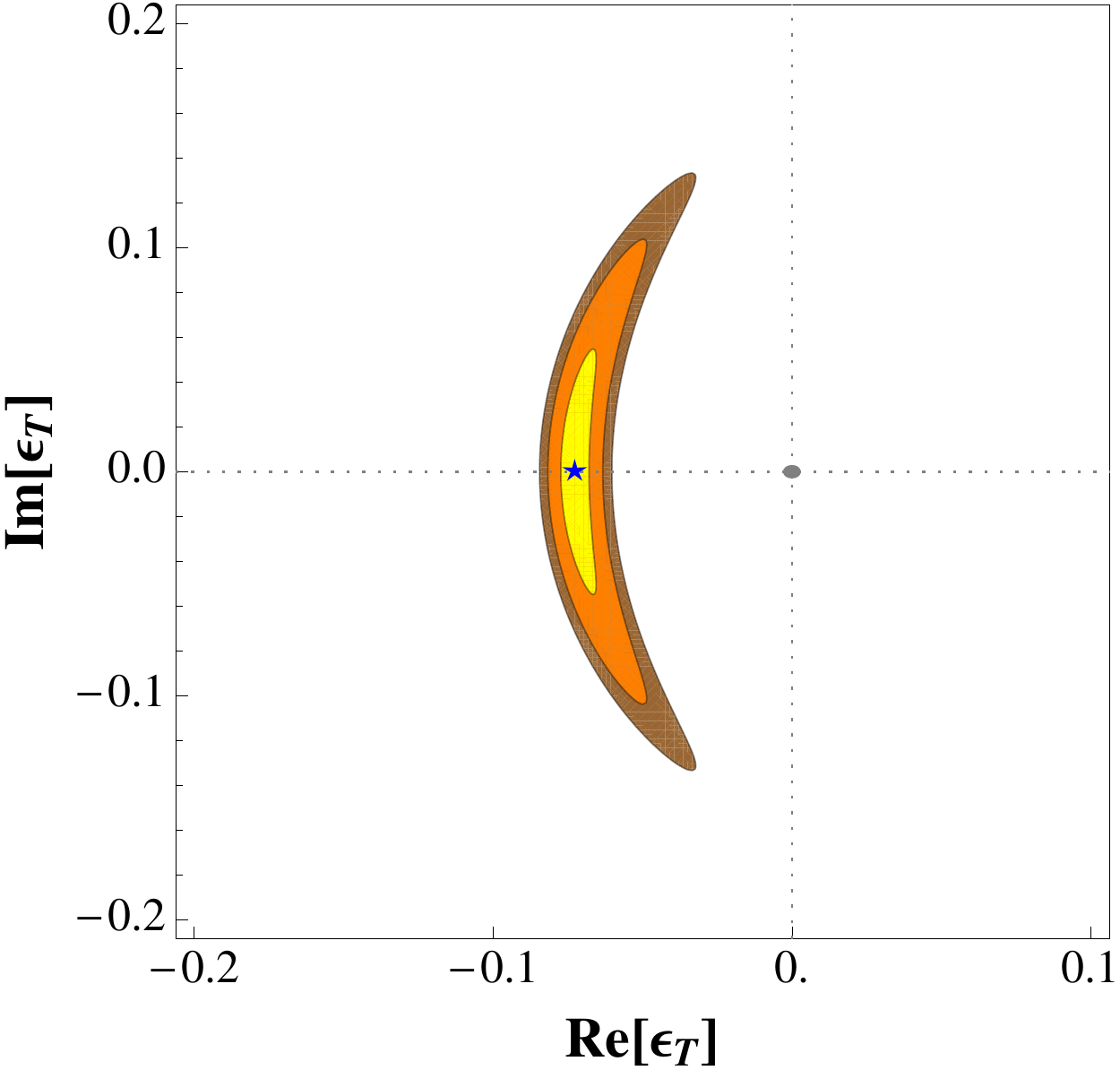}
\caption{(left)  Regions in the $(Re(\epsilon_T),Im(\epsilon_T))$ plane  determined from the experimental data (to $1$ and $2 \sigma$) on  ${\cal R}(D)$  (large rings) and ${\cal R}(D^*)$ (small rings).   (right)  Region  corresponding to values of $\chi^2$ between the minimum (indicated by the star), $1.55$ (yellow, light) and $1.65$ (orange, gray) and   $1.75$ (brown, dark).}\label{fig:oases}
\end{figure}

If  the   tensions in ${\cal R}(D)$ and ${\cal R}(D^*)$
are due to NP effects, it is interesting to investigate the  new  operator in the effective Hamiltonian (\ref{heff}) which  affects the observables in $B \to D^{(*)} \tau \nu_\tau$ transitions, focusing  on the  signatures  with  minimal dependence on hadronic quantities.  As done in \cite{Fajfer:2012vx,Fajfer:2012jt,Becirevic:2012jf,Datta:2012qk,Celis:2012dk,Crivellin:2012ye,Tanaka:2012nw},
${\cal R}(D)$ and ${\cal R}(D^*)$ data allow to constrain the values of the new effective dimensionless coupling.  In our case  $\epsilon_T$ is bounded   as shown in fig.\ref{fig:oases}.
Using the parameterization
\be
\epsilon_T=|a_T| e^{i \theta}+\epsilon_{T0} \,\,\, ,
 \ee
the tightest bound to $\epsilon_{T0}$ and $|a_T|$ is obtained from the measurement of ${\cal R}(D^*)$, while the combination of ${\cal R}(D)$ and ${\cal R}(D^*)$ data fixes the  range of the phase $\theta$.  We select the overlap of the two regions determined by ${\cal R}(D)$ and ${\cal R}(D^*)$ both at $1 \sigma$. In this overlap region, the function
$\chi^2(\epsilon_T)=\left( \frac{{\cal R}(D,\epsilon)-{\cal R}(D)^{exp}}{\Delta{\cal R}(D)^{exp}}\right)^2+\left( \frac{{\cal R}(D^*,\epsilon)-{\cal R}(D^*)^{exp}}{\Delta{\cal R}(D^*)^{exp}}\right)^2$ has  values running between $1.51$ and $1.75$.
This permitted  range of $\epsilon_T$ is represented as
\bea
Re[\epsilon_{T0}]&=&0.17 \,\,\,\,\, ,\,\, Im[\epsilon_{T0}]=0 \,\,\, ,  \nn \\
|a_T| &\in& [0.24,\,0.27]   \label{ranges-for-epsilonT}\\
\theta &\in&[2.6,\,3.7]\, {\rm rad}  \,\, \nn
\eea
and is also depicted in fig.\ref{fig:oases}.
Varying the effective coupling in this region we can analyze the impact of the new operator on various differential distributions.

We start with the longitudinal and transverse $D^*$ polarization distributions in $ B \to D^* \tau {\bar \nu}_\tau$.
\begin{figure}[!b]
\centering
\includegraphics[width = 0.4\textwidth]{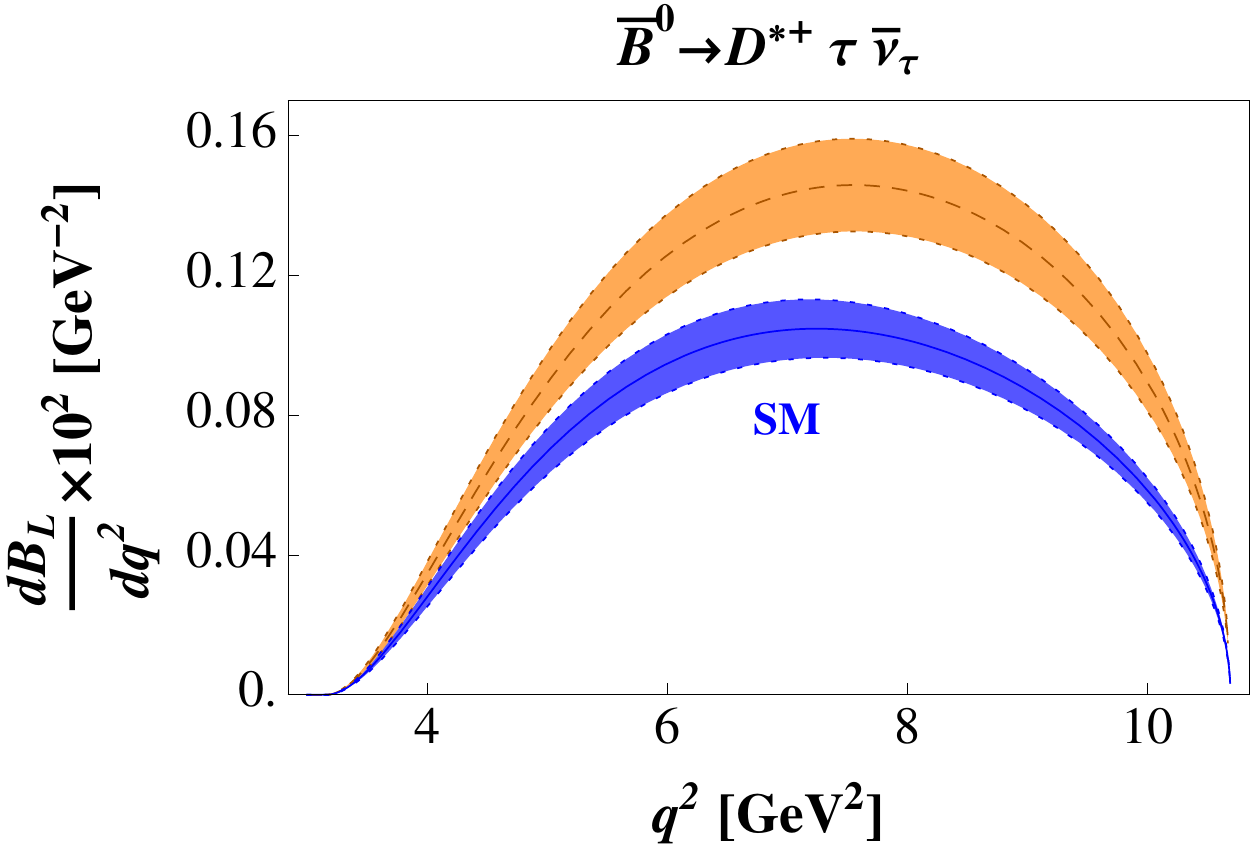}\hspace*{0.2cm}
\includegraphics[width = 0.4\textwidth]{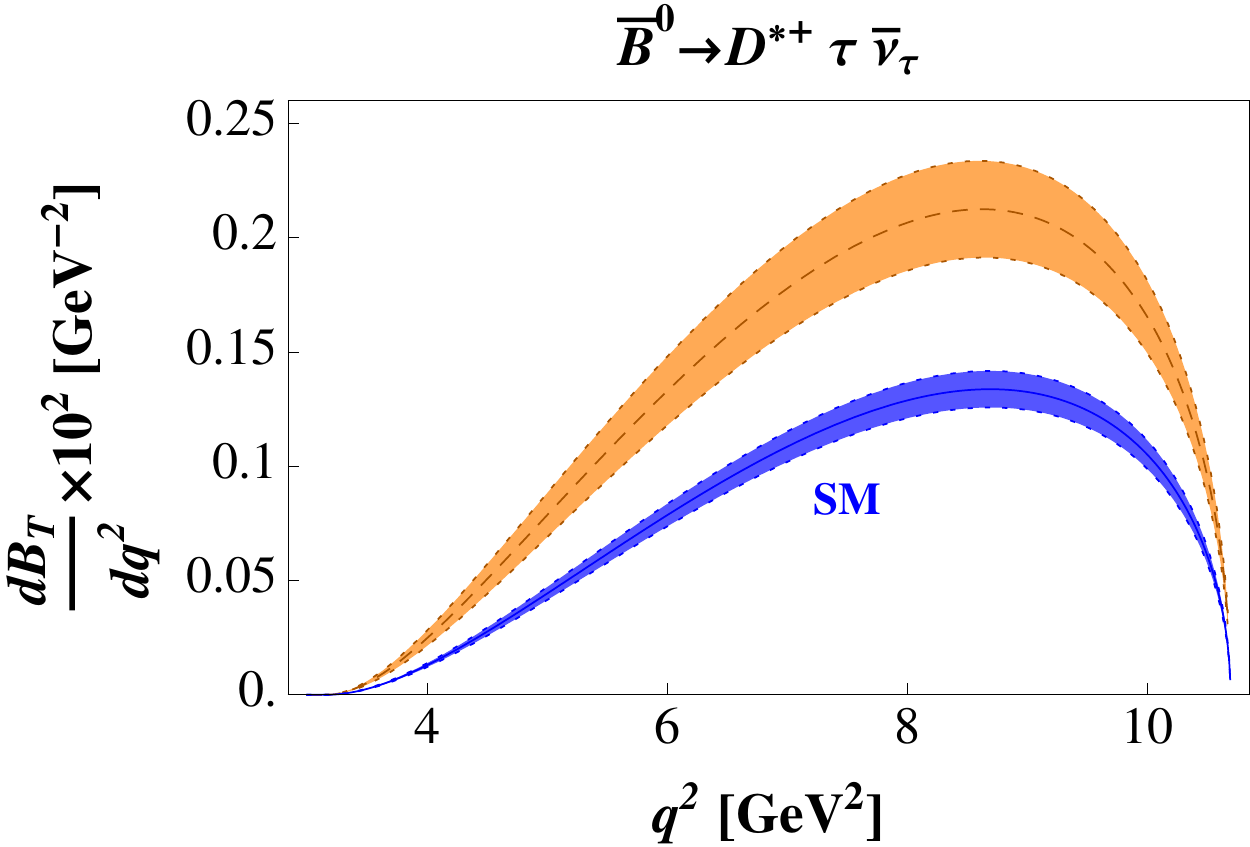}\\
\caption{
Differential branching ratios with polarized $D^*$:  $\displaystyle{\frac{d {\cal B}(B \to D^* \tau {\bar \nu}_\tau)_{L}}{dq^2}}$ (left) and $\displaystyle{\frac{d {\cal�B} (B \to D^* \tau {\bar \nu}_\tau)_T}{dq^2}}$ (right). The lower (blue) bands
are the SM prediction, the upper   (orange) bands include NP effects. In SM, the uncertainties on the parameters of the  Isgur-Wise function  in Eq.(\ref{belle-par}), together  with the errors on $\overline{\Lambda}$ and $\alpha_s$  are included. In the case of the  NP curves,   the uncertainty on $\epsilon_T$ is also  considered.}
\label{fig:dBrTL}
\end{figure}
We   consider the decay to a $D^*$ with definite helicity,  with  differential decay width  $\displaystyle{\frac{d \Gamma_{L,\pm}}{dq^2}}$ for the three cases  $L, \pm$. We  define $\displaystyle{\frac{d \Gamma_T}{dq^2}=\frac{d \Gamma_+}{dq^2}+\frac{d \Gamma_-}{dq^2}}$\,, and show
 in fig.\ref{fig:dBrTL} the differential branching fractions. The uncertainty in the distributions reflects the uncertainty on the parameters of the Belle Isgur-Wise function,  on $\bar \Lambda$ and, in the case of  NP,   on $\epsilon_T$.  While the shape of the distributions are slightly  modified from SM to NP,  the maxima increase, a consequence of  the increase of the branching fractions.

\begin{figure}[!b]
\centering
\includegraphics[width = 0.4\textwidth]{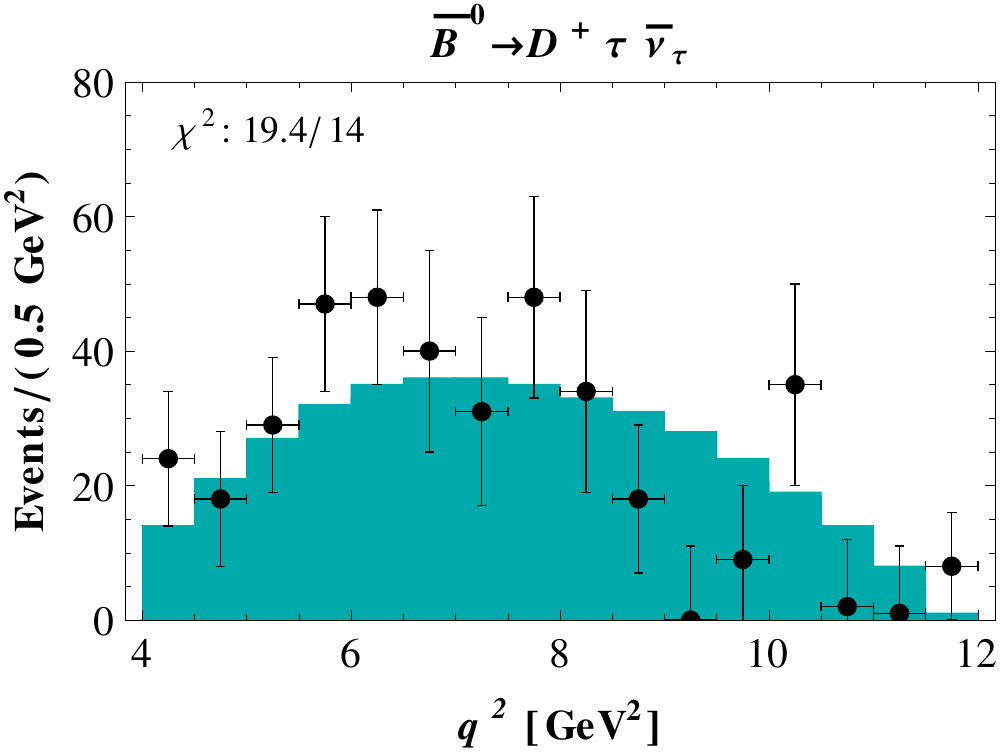}\hspace*{0.2cm}
\includegraphics[width = 0.4\textwidth]{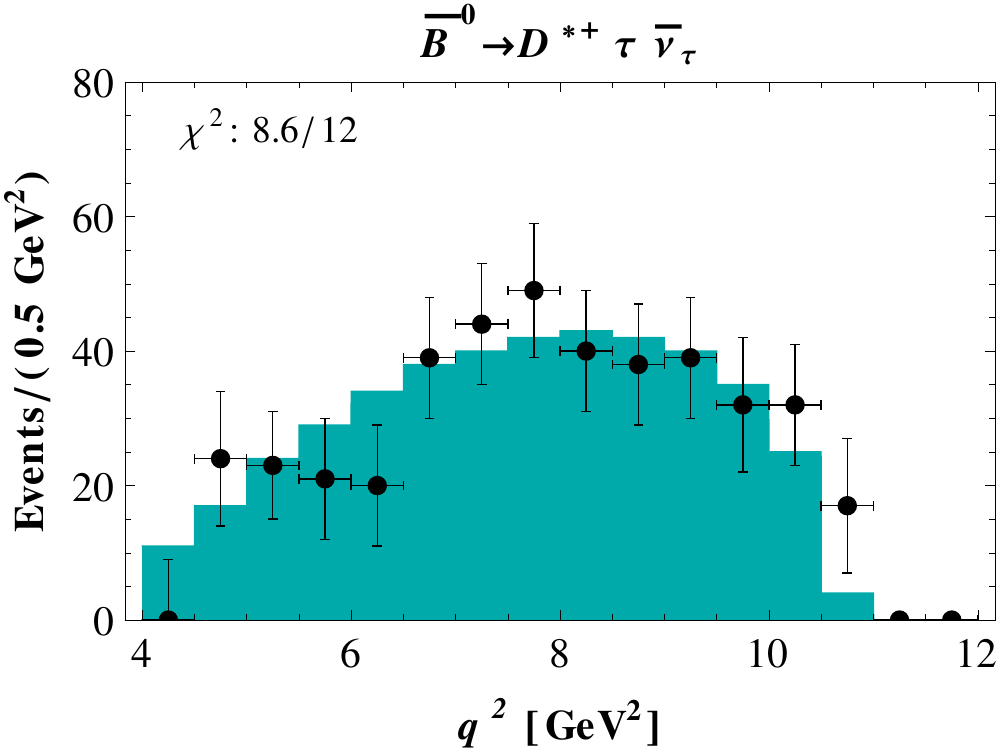}\\
\caption{
 $\displaystyle{\frac{d \Gamma(B \to D \tau {\bar \nu}_\tau)}{dq^2}}$ (left) and $\displaystyle{\frac{d \Gamma(B \to D^* \tau {\bar \nu}_\tau)}{dq^2}}$ (right) distributions in the NP scenario (for the central value of $\epsilon_T$, shaded histograms) compared to BaBar data (points) \cite{Lees:2013uzd}; the distributions are normalized to the total number of events.}
\label{fig:newspectrum}
\end{figure}

 The differential decay width distributions for $D$ and $D^*$ (summed over the $D^*$ polarizations) have been measured by BaBar
 \cite{Lees:2013uzd}, and can be compared to the SM and the NP scenario predictions. Once normalized to the total number of events, not only the SM distributions are compatible with data, as remarked in  \cite{Lees:2013uzd}, but also the distributions in the considered  NP scenario agree with measurements, as one can argue considering fig.\ref{fig:newspectrum}. This confirms that the shape of such distributions  does not allow at present to select between these possibilities, and other observables should be analyzed for a more efficient discrimination.

\begin{figure}[!h]
\centering
\includegraphics[width = 0.4\textwidth]{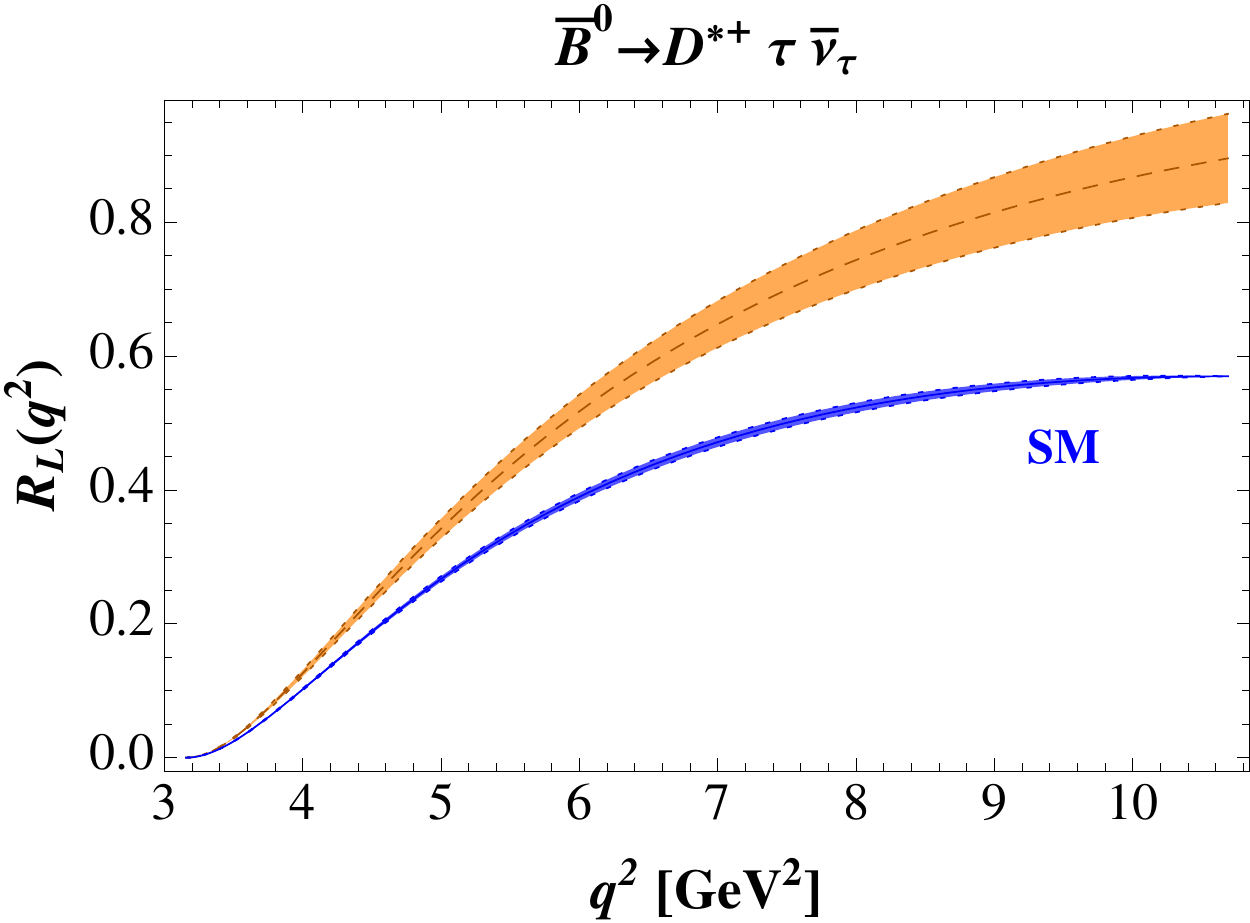}\hspace*{0.4cm}
\includegraphics[width = 0.4\textwidth]{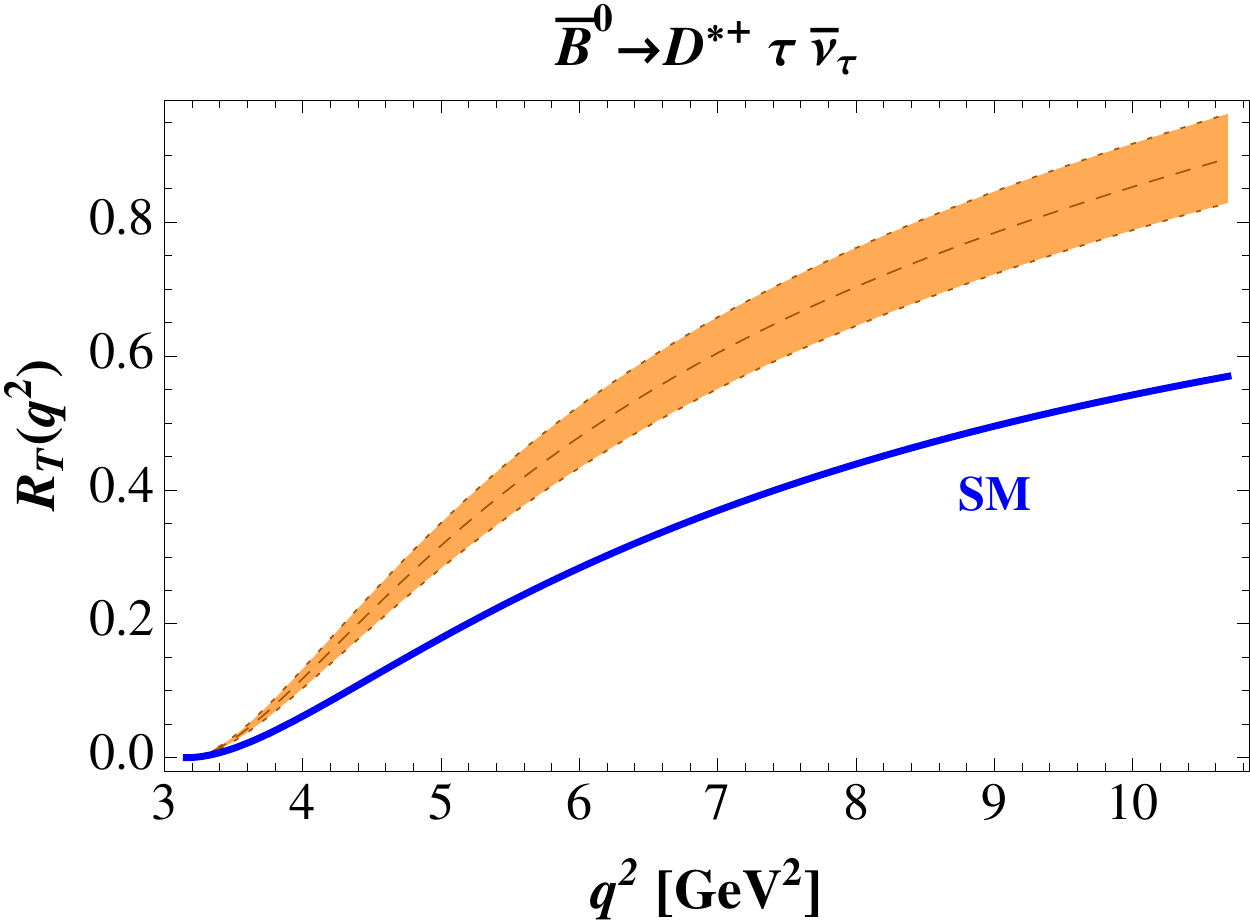}
\caption{$D^*$ polarization ratios $R_L^{D^*}(q^2)$ (left) and $R_T^{D^*}(q^2)$ (right) defined in (\ref{RLT}). Notations are the same as in fig.\ref{fig:dBrTL}.}\label{fig:RL}
\end{figure}

 \begin{figure}[!b]
\centering
\includegraphics[width = 0.4\textwidth]{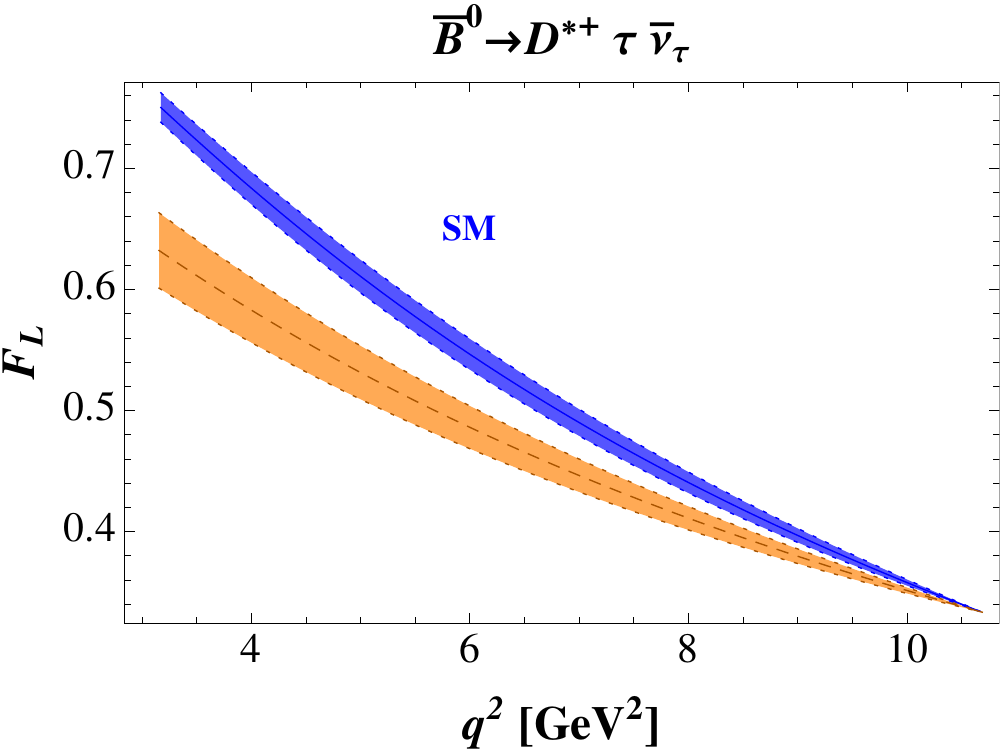}\hspace*{0.4cm}
\includegraphics[width = 0.4\textwidth]{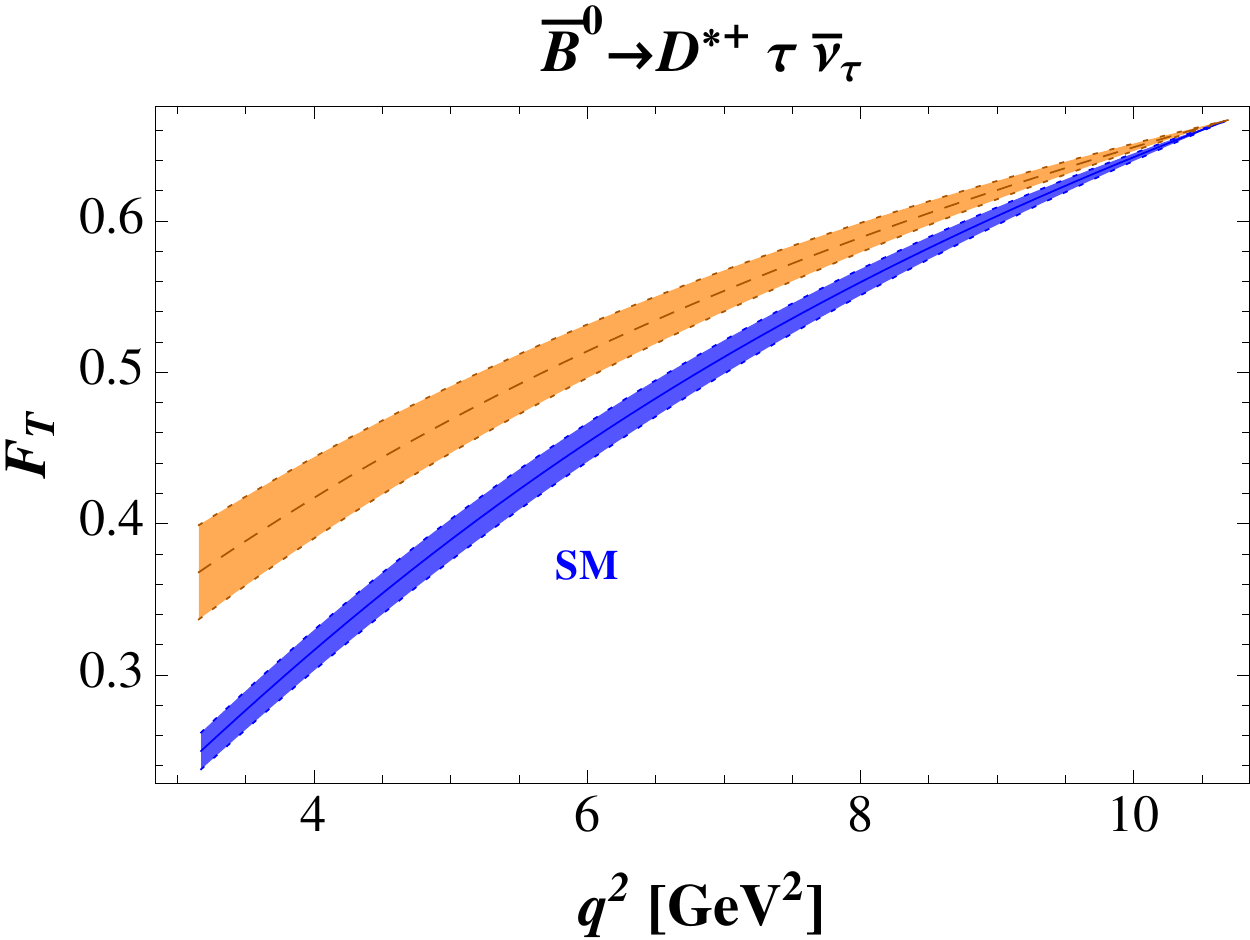}
\caption{ Polarization fractions $F_L(q^2)$ (left) and $F_T(q^2)$ (right) for  $B \to D^* \tau {\bar \nu}_\tau$ defined in (\ref{FLT}). Notations are the same as in fig.\ref{fig:dBrTL}.}\label{fig:ft}
\end{figure}

Other observables are the longitudinal and transverse $D^*$  polarization  distributions in $B \to D^* \tau {\bar \nu}_\tau$ normalized to $B \to D^* \ell {\bar \nu}_\ell$. They are   defined as
\be
R_ {L,T}^{D^*}(q^2)=\frac{d \Gamma_{L,T}(B \to D^* \tau {\bar \nu}_\tau)/dq^2}{d \Gamma_{L,T}(B \to D^* \ell {\bar \nu}_\ell)/dq^2} \,\,\, .  \label{RLT}
\ee
The SM predictions are shown in fig. \ref{fig:RL} together with  the modifications induced by the tensor  operator. At large $q^2$ the observables are enhanced by $30-50$ $\%$, a noticeable effect.
Furthermore, at odds with scenarios in which
only $R_L$ is affected by new physics    \cite{Celis:2012dk}, in the case of the tensor operator  both the longitudinal and the transverse  $R_L$ and $R_T$ distributions are modified.

The longitudinal and transverse polarization fractions of the $D^*$ meson
\be
F_{L,T}(q^2)=\frac{d \Gamma_{L,T}(B \to D^* \tau {\bar \nu}_\tau)}{dq^2} \times  \left(\frac{d \Gamma(B \to D^* \tau {\bar \nu}_\tau)}{dq^2}\right)^{-1} \label{FLT}
\ee
   are shown in fig.\ref{fig:ft}.
Both the SM and NP predictions are affected by a small error, since in the heavy quark limit the observables  in (\ref{FLT}) are  free of hadronic uncertainties,  due to the cancellation of  the form factor $\xi(w)$  in the ratio. The residual uncertainty reflects that on  $\bar \Lambda$ which controls the  $1/m_Q$ corrections.  The  uncertainty on $\bar \Lambda$ also enters in the  curves obtained in the NP scenario in combination with  $\epsilon_T$.
In  SM, $F_L(q^2)$ ranges between 0.75 at low $q^2$ and about 0.35 at high squared momentum transfer;  in  NP in the allowed region of $\epsilon_T$, $F_L(q^2)$ is between 0.35 and about 0.65 at low $q^2$, while this observable converges to the SM value at high $q^2$.
The SM predicts a dominant  longitudinal polarization  at small $q^2$,  in  NP the longitudinal and transverse polarizations have similar fractions up to $q^2=6$ GeV$^2$.

\begin{figure}[!b]
 \centering
\includegraphics[width = 0.4\textwidth]{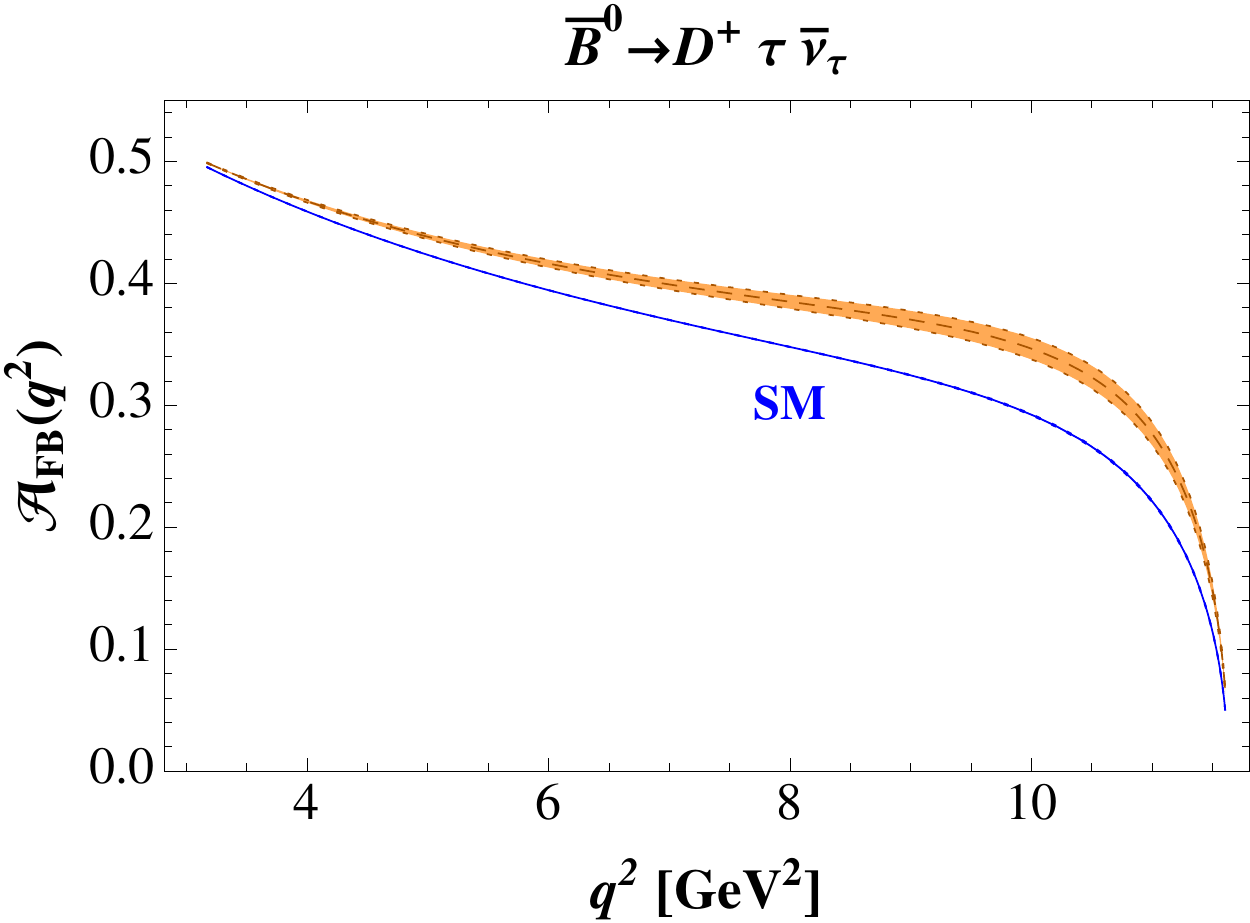}\hspace*{0.5cm}
\includegraphics[width = 0.4\textwidth]{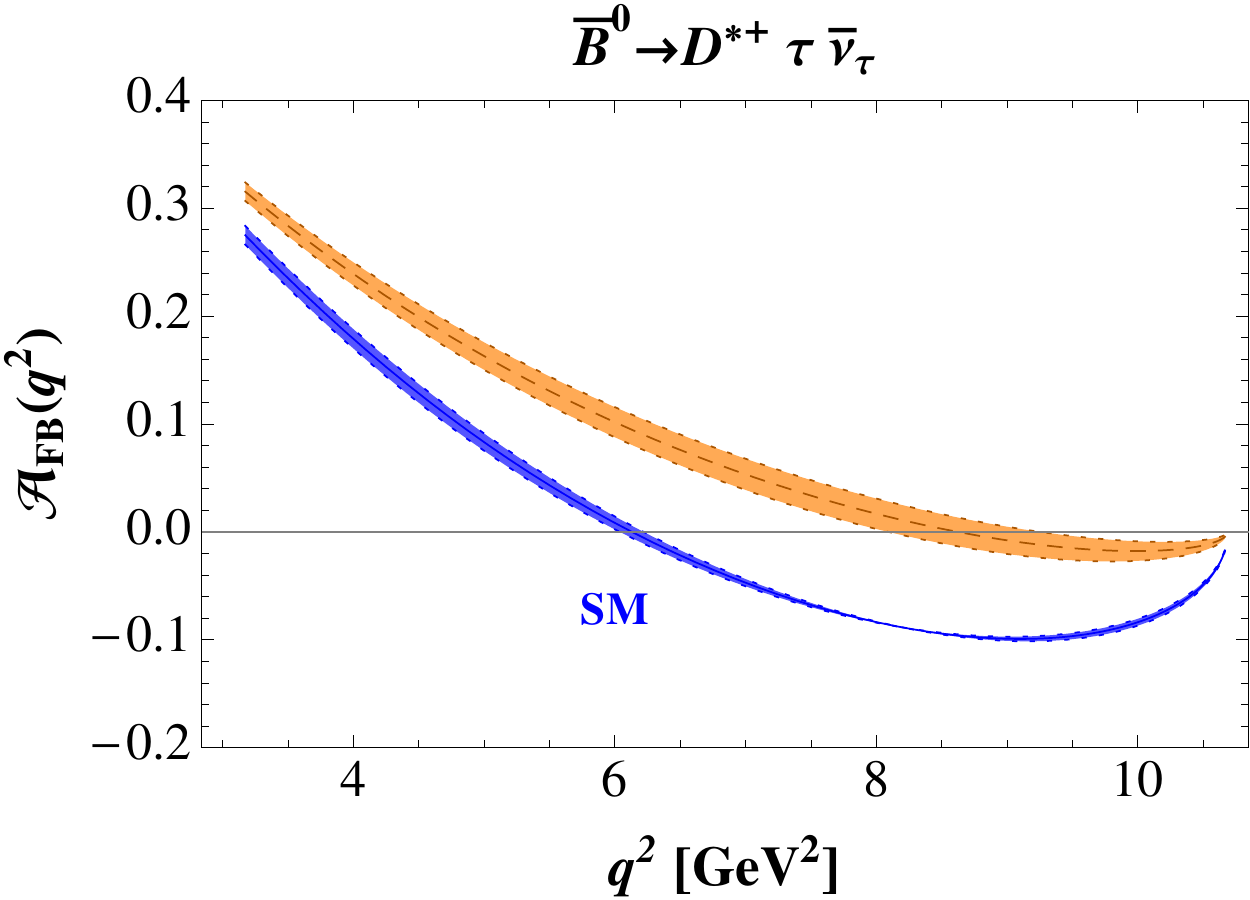}
\caption{ Forward-backward asymmetry ${\cal A}_{FB}(q^2)$ for  $B \to D \tau {\bar \nu}_\tau$ (left) and  $B \to D^* \tau {\bar \nu}_\tau$ (right). The lower (blue) curves are the SM predictions,  the upper (orange) bands  the NP expectations.
Uncertainty on $\overline{\Lambda}$ has been included and, in the case of  NP, also on the parameters $|a_T|$ and $\theta$. }\label{fig:afb}
\end{figure}

An important observable is the forward-backward  ${\cal A}_{FB}(q^2)$ asymmetry in  $ B \to D \tau {\bar \nu}_\tau$ and  $ B \to D^* \tau {\bar \nu}_\tau$,   defined as
\be
{\cal A}_{FB}(q^2)= \frac{\int_0^1 \,d \cos \theta_\ell \,\frac{d \Gamma}{dq^2 d \cos \theta_\ell} -\int_{-1}^0 \,d \cos \theta_\ell \, \frac{d \Gamma}{dq^2 d \cos \theta_\ell}}{\frac{d \Gamma}{dq^2}} \,\,\, ,
\label{eq:AFB}
\ee
where $\theta_\ell$ is the angle between the direction of the charged lepton  and the $D^{(*)}$ meson in the lepton pair rest frame.
We use the  notation
\be
{\cal A}_{FB}(q^2)= \frac{1}{\frac{d \Gamma}{dq^2}}\frac{3C(q^2)}{16}\left\{ \tilde{\cal  A}_{FB}^{SM}(q^2)+\tilde{\cal  A}_{FB}^{NP}(q^2)+\tilde{\cal A}_{FB}^{INT}(q^2) \right\} \,\,\, ,
\ee
with $C(q^2)$   defined in  (\ref{C-factor}) and the three terms in the parentheses given for $D$ and $D^*$:
\begin{itemize}
\item $D$
\bea
\tilde{\cal  A}_{FB}^{SM}(q^2)&=& 8F_0(q^2)F_1(q^2)(m_B^2-m_D^2)\frac{m_\ell^2}{q^2}\left(1 -\frac{m_\ell^2}{q^2}\right)\lambda^{1/2}(m_B^2,m^2,q^2)
\nn
\\
\tilde{\cal  A}_{FB}^{NP}(q^2)&=&0 \nn \\
\tilde{\cal  A}_{FB}^{INT}(q^2)&=& - 8 Re(\epsilon_T) F_0(q^2)[F_T(q^2)+G_T(q^2)] (m_B-m_D)m_\ell \left(1 -\frac{m_\ell^2}{q^2}\right)\lambda^{1/2}(m_B^2,m^2,q^2)
\eea
\item $D^*$
\bea
\tilde{\cal  A}_{FB}^{SM}(q^2)&=&\frac{4}{m_{D^*}(m_B+m_{D^*})q^2}\left(1 -\frac{m_\ell^2}{q^2}\right) \lambda^{1/2}(m_B^2,m_{D^*}^2,q^2) \nn \\
&&\Big\{m_\ell^2 A_0(q^2) \left[ A_1(q^2)(m_B+m_{D^*})^2(m_B^2-m_{D^*}^2-q^2)-\lambda(m_B^2,m_{D^*}^2,q^2)A_2(q^2) \right] \nn \\
&&-4m_{D^*}(m_B+m_{D^*})q^4A_1(q^2)V(q^2) \Big\}
\eea
\bea
\tilde{\cal  A}_{FB}^{NP}(q^2)&=& 16 |\epsilon_T|^2 \frac{m_\ell^2}{q^2}\left(1 -\frac{m_\ell^2}{q^2}\right) \lambda^{1/2}(m_B^2,m_{D^*}^2,q^2)({\tilde T}_1(q^2)+{\tilde T}_2(q^2))\nn \\
&&\left[(m_B^2-m_{D^*}^2)({\tilde T}_1(q^2)+{\tilde T}_2(q^2))+q^2({\tilde T}_1(q^2)-{\tilde T}_2(q^2)) \right]
\eea
\bea
\tilde{\cal  A}_{FB}^{INT}(q^2)&=& -4 Re(\epsilon_T) m_\ell \left(1 -\frac{m_\ell^2}{q^2}\right) \lambda^{1/2}(m_B^2,m_{D^*}^2,q^2) \Big\{ 4(m_B+m_{D^*}) A_1(q^2) ({\tilde T}_1(q^2)+{\tilde T}_2(q^2)) \nn \\
&+&A_0(q^2) \left[\frac{\lambda(m_B^2,m_{D^*}^2,q^2)}{m_{D^*}(m_B+m_{D^*})^2}{\tilde T}_0(q^2)+2\frac{m_B^2+m_{D^*}^2-q^2}{m_{D^*}}{\tilde T}_1(q^2)+4m_{D^*}{\tilde T}_2(q^2) \right] \nn \\
&-&\frac{V(q^2)}{m_B+m_{D^*}}\left[q^2( {\tilde T}_1(q^2)-{\tilde T}_2(q^2))+(m_B^2-m_{D^*}^2)({\tilde T}_1(q^2)+{\tilde T}_2(q^2)) \right]\Big\} \,\,\, .
\eea
\end{itemize}
In fig.\ref{fig:afb} we plot ${\cal A}_{FB}(q^2)$ for  $B \to D \tau {\bar \nu}_\tau$  and   $B \to D^* \tau {\bar \nu}_\tau$.
The SM prediction  is affected by almost no theoretical uncertainty, because of a nearly complete cancellation of the hadronic parameters   in the ratio.
In the case of   NP, we have taken into account also the uncertainty on  $\theta$ and $|a_T|$.
The SM curve lies in both cases below the NP distribution  for  all values of $q^2$. The most interesting deviation concerns  the $D^*$ mode: the SM predicts a zero for ${\cal A}_{FB}$  at  $q^2\simeq 6.15$ GeV$^2$,  in the NP case the zero is shifted towards larger values  $q^2 \in[8.1, 9.3]$ GeV$^2$.
Even though the experimental determination of the zero of the forward-backward asymmetry is challenging,  this observable  effectively  discriminates  SM from the NP model.
The integrated asymmetries, obtained integrating separately the numerator and the denominator in (\ref{eq:AFB}), are collected in Table \ref{tab:afb}:   for  $D^*$,    in the NP scenario  the integrated asymmetry has opposite  sign with respect to  SM.

\begin{table*}[!tb]
\centering
\begin{tabular}{|c | c |c |c|c|c| c|  }\hline
& ${\bar B}^0 \to D^+ \tau {\bar \nu}_\tau$ & ${\bar B}^0 \to D^{*+} \tau {\bar \nu}_\tau$ & ${\bar B}^0 \to D_0^{*+} \tau {\bar \nu}_\tau$ & ${\bar B}^0 \to D_1^{\prime +} \tau {\bar \nu}_\tau$ & ${\bar B}^0 \to D_1^+ \tau {\bar \nu}_\tau$ & ${\bar B}^0 \to D_2^{*+} \tau {\bar \nu}_\tau$ \\ \hline
${\cal A}_{FB}^{SM}$ & $0.357 \pm 0.002$ & $-0.040\pm 0.003$&$0.315$&$0.026$ & $ 0.24$ & $0.07$\\ \hline
${\cal A}_{FB}$  & $0.40 \pm 0.005$ &  $0.048 \pm 0.013$ &$0.30\pm0.005$&$0.08\pm0.01$& $0.21 \pm 0.003$ & $0.14 \pm 0.01$\\ \hline
 \end{tabular}
\caption{ Integrated forward-backward asymmetry for the   considered decay modes.
The first line reports  the SM results,  in the second line the effect of the  tensor operator is included.  }\label{tab:afb}
\end{table*}

\section{Tensor operator in $B \to D^{**} \ell {\bar \nu}_\ell$ decays}

The new operator in the effective hamiltonian (\ref{heff}) affects  other exclusive decay modes  that are worth investigating. Of peculiar interest are the semileptonic $B$ and $B_s$ transitions into  excited
charmed mesons. The lightest multiplet of such hadrons,  corresponding to the quark model  $p$-wave ($\ell=1$) mesons and generically denoted  $D_{(s)}^{**}$, comprises four positive parity states which, in the heavy quark limit, fill two  doublets labeled by the  (conserved) angular momentum ${\vec s}_\ell={\vec s}_q+{\vec \ell}$ (${\vec s}_q$ is  spin of  the light antiquark),  hence
 $s_\ell=1/2$ or $s_\ell=3/2$.
The two mesons belonging to the first doublet,  $(D^*_{(s)0},\,D_{(s)1}^\prime)$,  have spin-parity $J^P=(0^+,1^+)$;  the mesons in the second doublet have $J^P=(1^+,2^+)$ and are named $(D_{(s)1},\,D_{(s)2}^*)$. All the members of the doublets, with and without strangeness,  have been observed, and the two  $s_\ell^P=1/2^+$  states without strangeness are found to be broad, as expected \cite{Colangelo:2012xi}.

In the heavy quark limit also the semileptonic $B$ transitions to mesons belonging to the same charmed doublet can be described in terms of a single form factor.   $B$ decays to $(D^*_0,\,D_1^\prime)$ are governed by a  universal function denoted as $\tau_{1/2}(w)$,  $B$ decays to $(D_1,\,D_2^*)$  by the  $\tau_{3/2}(w)$ form factor (the matrix elements  are collected  in appendix  \ref{app:me}).
There are several determinations of the $\tau_i(w)$ parametrized in terms of the zero-recoil value $\tau_i(1)$ (contrary to the Isgur-Wise function, $\tau_i(w)$ are not normalized to unity at  $w=1$), of the slope $\rho_i^2$ and of the curvature $c_i$. In the ratios of branching fractions and asymmetries  the zero-recoil value does not play any role, and this reduces the main dependence of the observables on the hadronic parameters. The present experimental situation needs to be settled, since the semileptonic $B$ decay rates into  $(D^*_0,\,D_1^\prime)$ exceed the predictions obtained using computed $\tau_i(1)$;  the origin of the discrepancy is still unknown, and could be related to the broad widths of the final charmed mesons,  which determine a difficulty in the exclusive reconstruction,  and to a possible pollution from other (e.g. radial) excited states. Semileptonic  $B_s$ decays to $s_\ell^P=1/2^+$ $c \bar s$ mesons could clarify the issue, due to the  narrow width of the strange charmed resonances
 \cite{Becirevic:2012te}. On the other hand,  the  tensor operator   produces precise correlations among various observables, therefore its effects could  be distinguished from others.

 For definiteness,   we use a QCD sum rule
determination of $\tau_{3/2}(w)$   at leading order in  $\alpha_s$   \cite{Colangelo:1992kc,Colangelo:1998sf}, and of  $\tau_{1/2}(w)$   at ${\cal O}(\alpha_s)$  \cite{Colangelo:1998ga}:
\bea
\tau_{3/2}(w)&=&\tau_{3/2}(1) \left[1-\rho^2_{3/2}(w-1) \right] \label{tau32}
\\
\tau_{1/2}(w)&=&\tau_{1/2}(1) \left[1-\rho^2_{1/2}(w-1) +c_{1/2}(w-1)^2 \right] \label{tau12}
\eea
with
\bea
\tau_{3/2}(1)&=& 0.28 \hskip 2. cm \rho^2_{3/2}=0.9 \label{32parameters}
\\
\tau_{1/2}(1)&=& 0.35 \pm 0.08 \hskip 1 cm \rho^2_{1/2}=2.5 \pm 1.0  \hskip 1 cm c_{1/2}=3 \pm 3 \,\,.\label{12parameters}
\eea
The  differential decay rates for  $B \to D^{**} \ell {\bar \nu}_\ell$  can be written as in (\ref{dgammadq2-generic}), see   appendix  \ref{app:me}.
The ratios
\be
{\cal R}(D^{*}_0)=\frac{{\cal B}(B \to D^{*}_0 \tau \,{\bar \nu}_\tau)}{{\cal B}(B \to D^{*}_0 \ell \,{\bar \nu}_\ell)}
\label{rdstarstar}
\ee
and the analogous ${\cal R}(D^\prime_1)$, ${\cal R}(D_1)$ and ${\cal R}(D^*_2)$ depend on the effective coupling  $\epsilon_T$.  This also happens in   $B_s \to D^{**}_s \ell {\bar \nu}_\ell$ transitions, in the  $SU(3)_F$ symmetry limit for the form factors.

\begin{figure}[!h]
 \centering
\includegraphics[width = 0.35\textwidth]{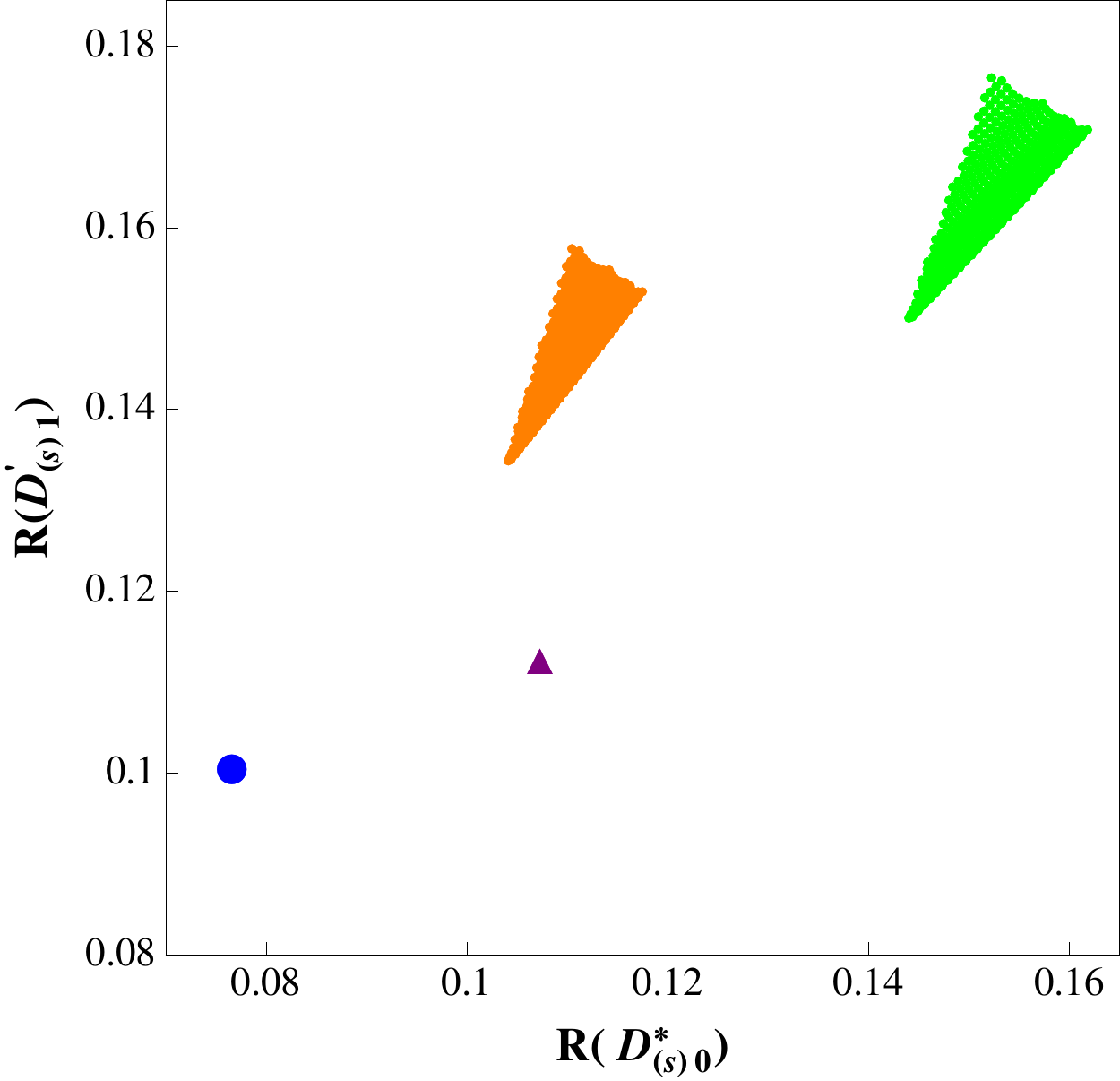}\hspace*{0.5cm}
\includegraphics[width = 0.35\textwidth]{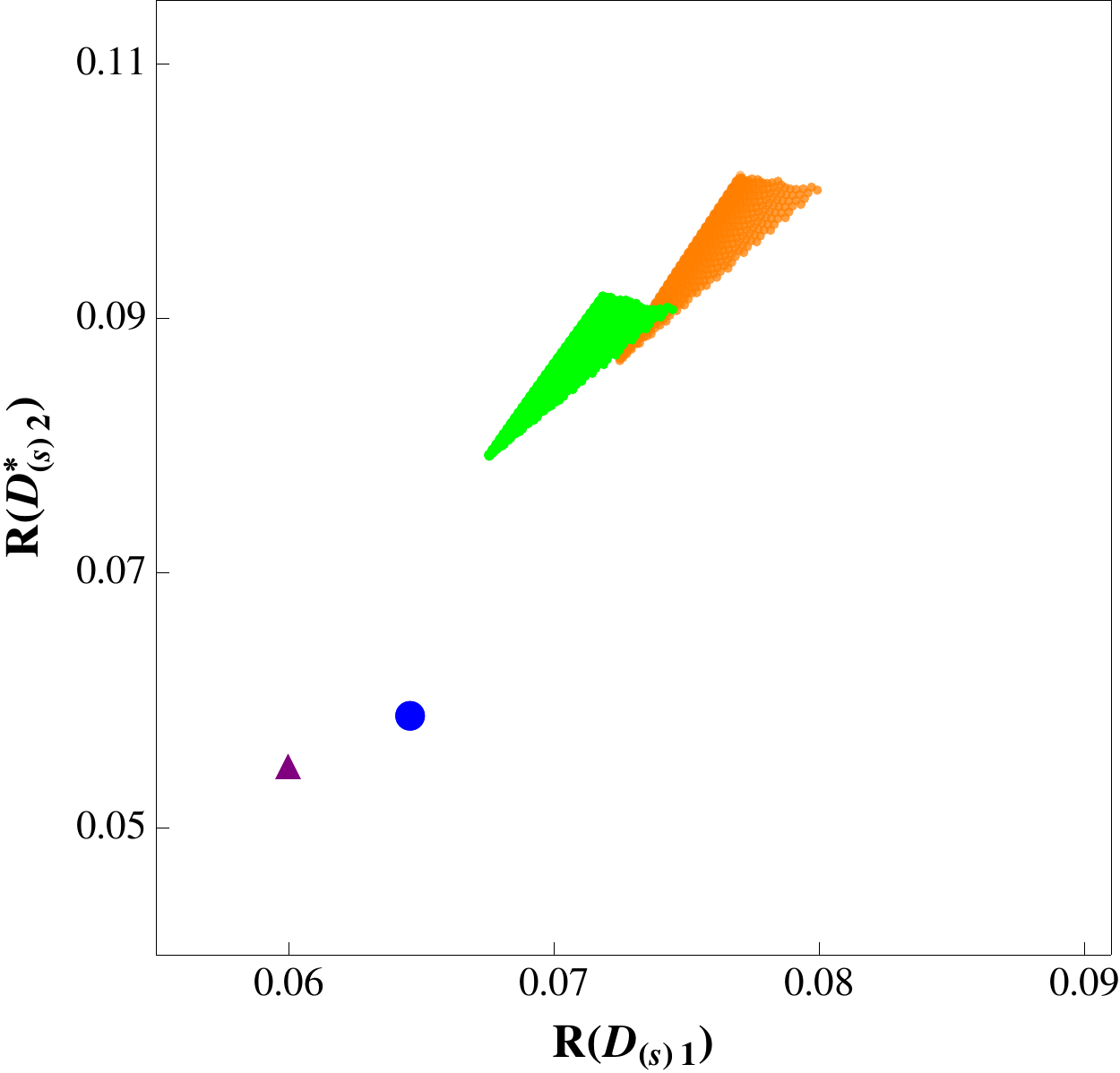}
\caption{(left) Correlations between the ratios  ${\cal R}(D^{*}_{(s)0})$ and ${\cal R}(D^\prime_{(s)1})$ for  mesons belonging to the  $(D^*_{(s)0},\,D_{(s)1}^\prime)$ doublet without (orange, dark) and with strangeness (green, light).  (right) Correlation between  ${\cal R}(D_{(s)1})$ and ${\cal R}(D^*_{(s)2})$  for  mesons in  the   $(D_{(s)1},\,D_{(s)2}^*)$ doublet. The  dots  (triangles) correspond to the SM results for mesons  without (with) strangeness.
}\label{fig:r12-r32}
\end{figure}

In fig.\ref{fig:r12-r32}   for each meson doublet  we show the correlation between the ratios   (\ref{rdstarstar}) for    $B$ and  $B_s$, together with
the  SM predictions $({\cal R}(D^{*}_{0}),{\cal R}(D^\prime_{1}))=(0.077,0.100)$, $({\cal R}(D^{*}_{s0}),{\cal R}(D^\prime_{s1}))=(0.107,0.112)$, $({\cal R}(D_{1}),{\cal R}(D^*_{2})=(0.065,0.059)$ and $({\cal R}(D_{s1}),{\cal R}(D^*_{s2})=(0.060,0.055)$.
The tensor operator  produces a sizeable increase in the  ratios $\cal R$,  which is correlated for the two members in  each doublet.
The hadronic uncertainty is  mild:  using  the $\tau_i$ functions in \cite{Morenas:1997nk}, the results remain almost unchanged in the case of the $s_\ell=3/2$ doublet, while  for  $s_\ell=1/2$  they are smaller by about $25\%$  in  SM and in the NP case. The same effect is found using the form factors  obtained by lattice QCD \cite{Blossier:2009vy}.

\begin{figure}[!h]
 \centering
 \includegraphics[width = 0.4\textwidth]{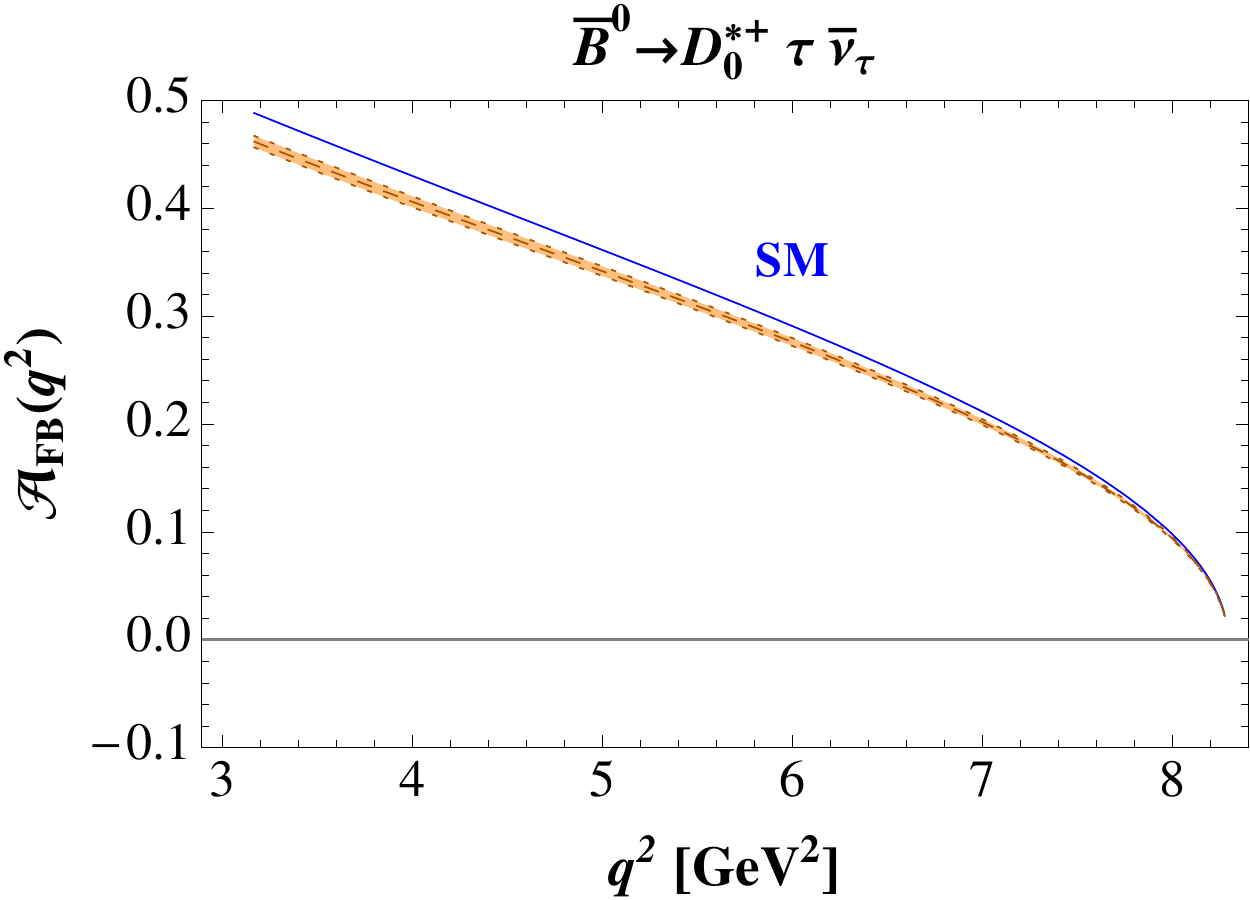}\hspace*{0.5cm}
\includegraphics[width = 0.4\textwidth]{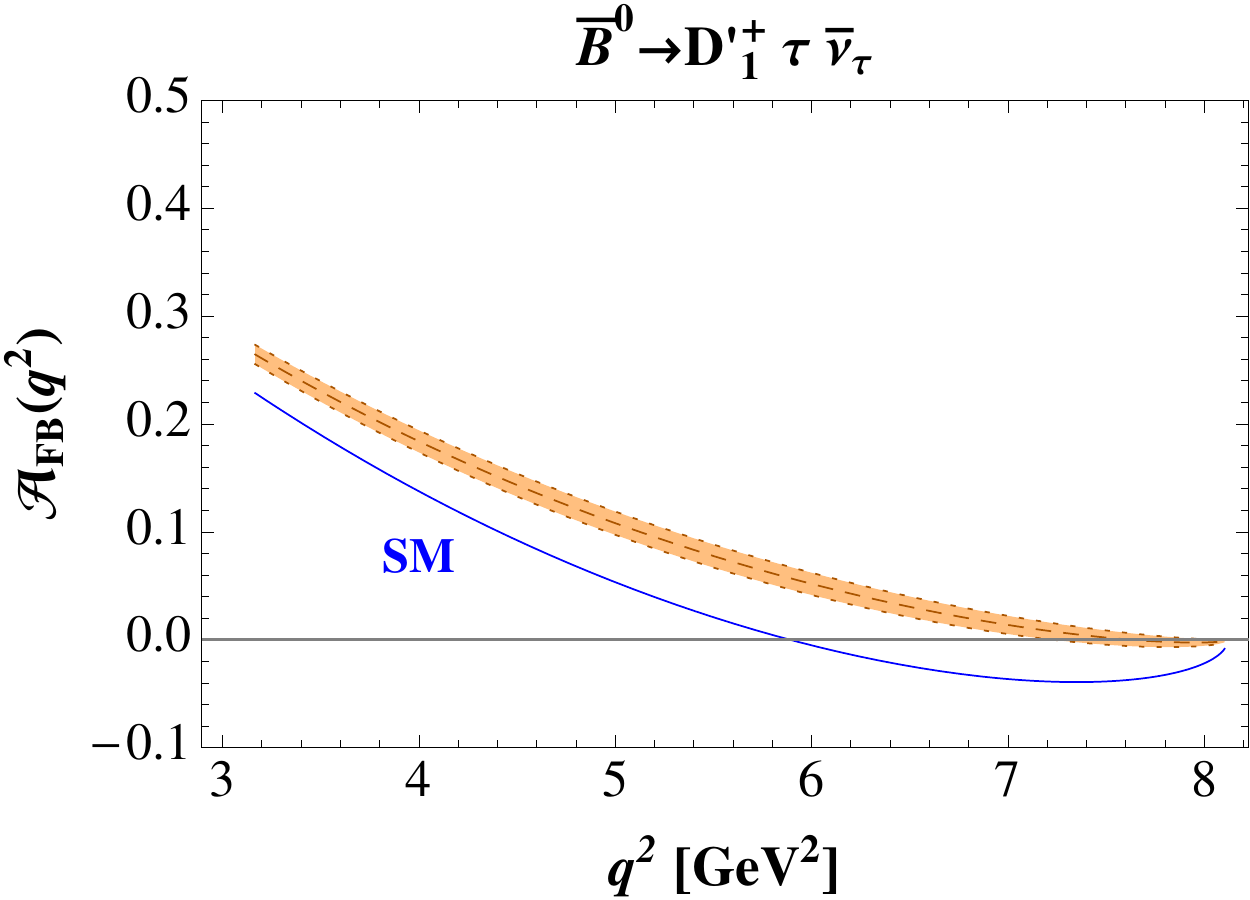}\\\vspace*{0.2cm}
\includegraphics[width = 0.4\textwidth]{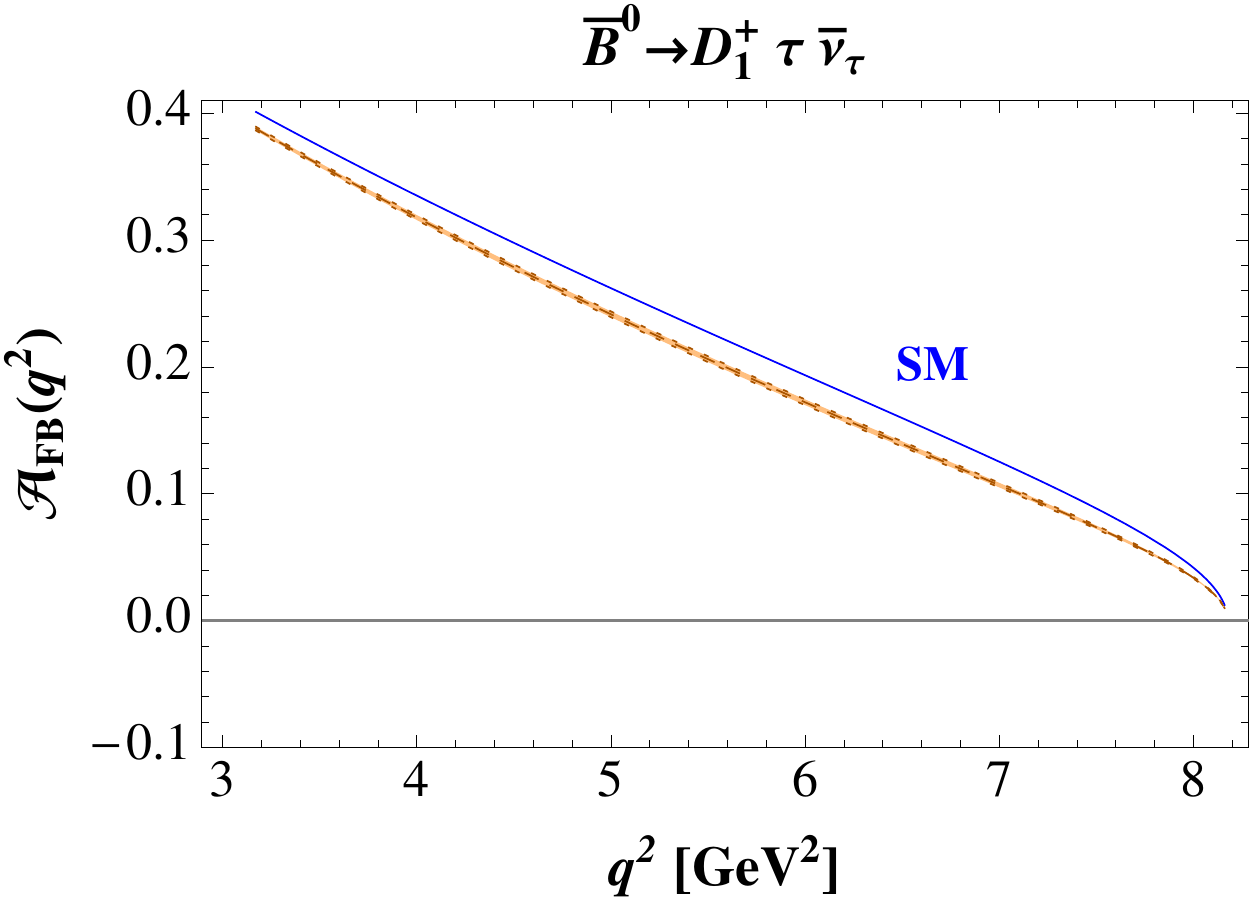}\hspace*{0.5cm}
\includegraphics[width = 0.4\textwidth]{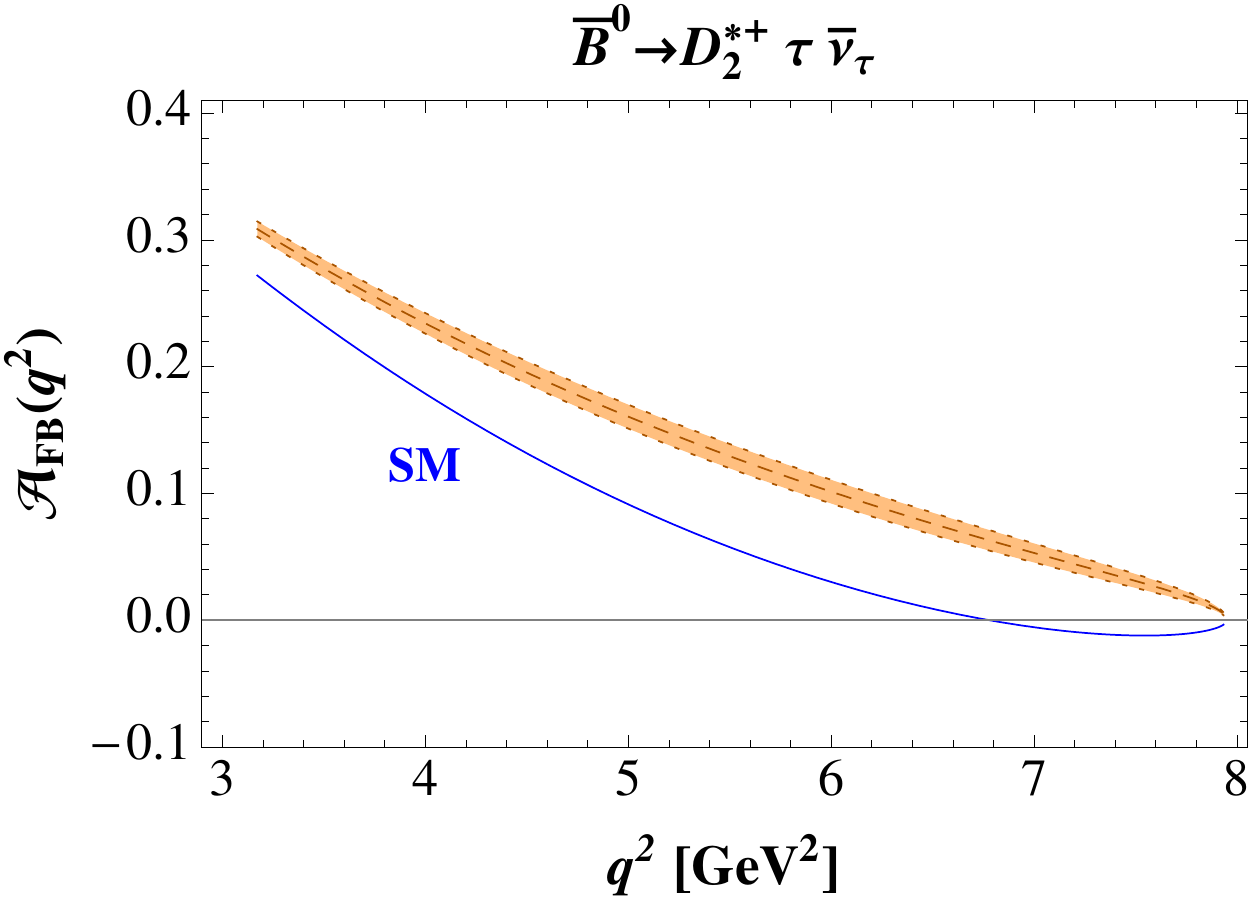}
\caption{Forward-backward asymmetry ${\cal A}_{FB}$ for the decays $B \to D_0^* \tau {\bar \nu}_\tau$ (top, left), $B \to D^\prime_1 \tau {\bar \nu}_\tau$ (top, right),
 $B \to D_1 \tau {\bar \nu}_\tau$ (bottom, left) and  $B \to D^*_2 \tau {\bar \nu}_\tau$ bottom, (right)  as function of $q^2$. The solid (blue) curves are the SM predictions,  the dotted (orange) bands  the NP expectations. }\label{fig:afbD**}
\end{figure}

The differential forward-backward asymmetries in the case of the four positive parity charmed mesons are collected in fig.\ref{fig:afbD**}, and the integrated ones in Table \ref{tab:afb}.
While in  $B \to (D^*_0, D_1) \tau  {\bar \nu}_\tau$ the forward-backward  asymmetry does not discriminate between SN and  NP, in the  modes with $D_1^\prime$ and
$D_2^*$ it is a sensitive observable: The inclusion of the tensor operator produces an enhancement of ${\cal A}_{FB}$ with respect to  SM  for all values of $q^2$. Moreover,  in SM there is  a zero
which,  in the case of $B \to D^\prime_1 \tau {\bar \nu}_\tau$ moves towards larger values of $q^2$, and  disappears in $B \to D^*_2 \tau {\bar \nu}_\tau$  once  NP is included.

We close this section remarking that, while  the tensor operator in (\ref{heff})  does not affect the purely leptonic $B_c \to \tau^- {\bar \nu}_\tau$ mode,    it can have an impact on the transitions $B_c \to (\eta_c, J/\psi)  \tau^- {\bar \nu}_\tau$ and
$\Lambda_b \to \Lambda_c \tau^- {\bar \nu}_\tau$; therefore,  sets  of other observables can be identified and investigated, with precise correlated deviations from the SM predictions.

\section{Conclusions}
The detailed experimental information  provided us  on  flavour physics shows  an astonishing consistency with the SM predictions. The very few tensions  identify  possible paths to new physics searches.
The BaBar anomalous enhancement  of the ratios $\displaystyle R(D^{(*)})=\frac{{\cal B}(B \to D^{(*)} \tau {\bar \nu}_\tau)}{{\cal B}(B \to D^{(*)} \mu {\bar \nu}_\mu)}$  with respect to SM is one of these few cases. The analyses  of  $R(D^{(*)})$ in specific models also evidentiate the enhancement the purely leptonic  $B \to \tau {\bar \nu}_\tau$ rate, for which  data are better compatible with SM.  A mechanism enhancing  the semileptonic modes  $B \to D^{(*)} \tau {\bar \nu}_\tau$ with respect to $B \to D^{(*)} \mu {\bar \nu}_\mu$, leaving $B \to \tau {\bar \nu}_\tau$ unaffected, can be based on a
 tensor operator in the effective hamiltonian. We have  bound  the relative weight $\epsilon_T$ of this  operator and studied the impact on several  observables, the most sensitive one being
the forward-backward asymmetry  in  $ B \to D^* \tau {\bar \nu}_\tau$ with a shift in  the position of its zero.
If the anomaly in      $B \to D^{(*)} \tau {\bar \nu}_\tau$ is due to this NP effect,  analogous deviations  should be found in B   to excited $D$  transitions. The ratios  $R$ for these mesons are enhanced with respect to SM, and the  forward-backward asymmetry is  a sensitive observable  in the channels involving $D_1^\prime$ and $D_2^*$.
These signatures in  exclusive semileptonic $b \to c \,\tau {\bar \nu}_\tau$ modes   make   the understanding of  the role of  the new contribution to the effective  weak hamiltonian feasible,  a step towards  possibly disclosing new interactions through flavour physics measurements.

\section*{Acknowledgement}
This work is supported in part by the Italian MIUR Prin 2009.

\appendix

\section{Coefficients}\label{app:coefficients}
With the aim of providing the information useful to reconstruct the various $B \to D^{(*)}$  matrix elements,  we collect here the expressions  of the $\alpha_s$ and $1/m_Q$ corrections in Eqs.(\ref{hpiu},\ref{hmeno}) and (\ref{hv}-\ref{ha3})   worked out by M. Neubert and by I. Caprini et al. in \cite{hqet,Caprini:1997mu}.  The functions $L_i(w)$ read as
\bea
L_1 &\simeq& 0.72 \, (w-1) \, {\bar \Lambda} \nn \\
L_2 &\simeq& -0.16 \, (w-1) \, {\bar \Lambda} \nn \\
L_3 &\simeq& -0.24 \, {\bar \Lambda} \nn \\
L_4 &\simeq& 0.24 \, {\bar \Lambda}\\
L_5 &\simeq& - {\bar \Lambda} \nn \\
L_6 &\simeq& - {3.24 \over w+1} \, {\bar \Lambda} \,\,\, .  \nn
\eea
The coefficients $C_i$  are expressed in terms of $C_1$,
\bea
{C_1^5 \over C_1}&=& 1-{4 \alpha_s \over 3 \pi} r_f(w) \nn \\
{C_2^{(5)}\over C_1}&=&-{2 \alpha_s \over 3 \pi} H_{(5)}\left(w,\frac{1}{z_m}\right) \label{app:a2} \\
{C_3^{(5)}\over C_1}&=&\mp {2 \alpha_s \over 3 \pi} H_{(5)}(w,z_m)  \,\,\, ,\nn
\eea
with
$z_m={m_c \over m_b}$ and
\bea
r_f(w)&=& {1 \over \sqrt{w^2-1}}\log \left[w+\sqrt{w^2-1}\right]  \,\,\,\, ,  \\
H_{(5)}(w,z_m) &=& {z_m (1-\log z_m \mp z_m) \over 1 -2w z_m +z_m^2}  +{z_m \over (1 -2w z_m +z_m^2)^2} \Big[2(w\mp 1)z_m(1\pm z_m) \log z_m \nn \\
&-&[(w\pm1)-2w(2w\pm 1)z_m+(5w\pm 2w^2\mp 1)z_m^2-2z_m^3]r_f(w) \Big] \,\,\, .\label{app:a4}
\eea
In (\ref{app:a2},\ref{app:a4}) the lower signs refer to the index $5$ (corresponding to the axial current).
$C_1$ reads:
\be
C_1=\left({ \alpha_s (m_c) \over \alpha_s (\mu)} \right)^{a_{hh}(w)} \left(1 - {\alpha_s (\mu) \over \pi} Z_{hh}(w)\right) \left(1 +
   {\alpha_s (m_c)\over \pi} \left[\log \left(\frac{m_b}{m_c}\right) + Z_{hh}(w) + {2 \over 3} [f(w) + r_f(w) + g(w)] \right] \right) \,\,\, ,
\ee
with
\bea
a_{hh}(w) &=& {8 \over 27} [w \, r_f(w) -1] \,\,\, ,  \\
Z_{hh}(w) &=&{8 \over 81} \left( {94 \over 9} - \pi^2 \right) (w-1)-{4 \over 135} \left( {92 \over 9} -\pi^2 \right) (w-1)^2 + {\cal O}((w-1)^3)\,\,\, ,  \\
f(w) &=& w r_f(w)-2 -{w \over \sqrt{w^2 -1}}[L_2(1- w_-^2)+(w^2-1)r_f^2(w)] \,\,\,\, ,  \\
g(w) &=& {w \over \sqrt{w^2 -1}}[L_2(1-z_m w_-)-L_2(1-z_m w_+)]- { z_m\over (1 -2w z_m +z_m^2)} [(w^2-1) r_f(w) +(w-z_m)\log (z_m)] \,\,\, , \nn \\
\eea
and $w_\pm= w \pm \sqrt{w^2 -1}$.
In the numerical analysis we set the scale  $\mu=\sqrt{m_c \,m_b}$, and investigate the sensitivity to higher order corrections  varying this scale between $\mu/2$ and $2 \mu$.

\section{$B \to D^{**}$ matrix elements and differential semileptonic decay rates}\label{app:me}
In the infinite heavy quark mass limit the $B \to D^{**}$ matrix elements can be defined in terms of two universal $\tau_{1/2}(w)$ and $\tau_{3/2}(w)$ form factors:
\bea
<D^*_0(p^\prime)|{\bar c} \gamma_\mu(1-\gamma_5) b| B(p)>&=&-2\sqrt{m_B m_{D^*_0}} \, \tau_{1/2}(w)\, \left(v-v^\prime \right)_\mu  \,\,\, ; \label{V-A-D0} \\
<D^*_0(p^\prime)|{\bar c} \sigma_{\mu \nu}(1-\gamma_5) b| B(p)>&=&2\sqrt{m_B m_{D^*_0}} \, \tau_{1/2}(w)\, \left[ -\epsilon_{\mu \nu \alpha \beta} v^\alpha v^{\prime \beta}+i(v_\mu v^\prime_\nu- v_\nu v^\prime_\mu) \right]  \,\,\, ; \label{T-D0}\\ \nn \\
<D^\prime_1(p^\prime,\,\epsilon)|{\bar c} \gamma_\mu(1-\gamma_5) b| B(p)>&=&-2\sqrt{m_B m_{D^\prime_1}} \, \tau_{1/2}(w)\, \left[-i \epsilon_{\mu \alpha \beta \sigma}\epsilon^{*\alpha}v^\beta v^{\prime \sigma}-(w-1)\epsilon^*_\mu+(\epsilon^* \cdot v)v^\prime_\mu \right]  \,\,\, ;  \label{V-A-D1primo} \\
<D^\prime_1(p^\prime,\,\epsilon)|{\bar c} \sigma_{\mu \nu}(1-\gamma_5) b| B(p)>&=&-2\sqrt{m_B m_{D^\prime_1}} \, \tau_{1/2}(w)\, \left\{ -\epsilon_{\mu \nu \alpha \beta} \epsilon^{*\alpha}(v- v^{\prime})^\beta+i[\epsilon^*_\mu (v-v^\prime)_\nu- \epsilon^*_\nu (v-v^\prime)_\mu] \right\} \,\,\, ; \nn \\  \label{T-D1primo} \\  \nn \\
<D_1(p^\prime,\,\epsilon)|{\bar c} \gamma_\mu(1-\gamma_5) b| B(p)>&=&\frac{\sqrt{m_B m_{D_1}}}{\sqrt{2}} \, \tau_{3/2}(w) \, \Big\{
i(1+w) \epsilon_{\mu  \alpha \beta \sigma}\epsilon^{*\alpha}v^\beta v^{\prime \sigma}+(w^2-1)\epsilon^*_\mu \nn \\
&+&(\epsilon^* \cdot v)\left[3v_\mu-(w-2)v^\prime_\mu \right] \Big\} \,\,\, ; \label{V-A-D1}  \\
<D_1(p^\prime,\,\epsilon)|{\bar c} \sigma_{\mu \nu}(1-\gamma_5) b| B(p)>&=&\frac{\sqrt{m_B m_{D_1}}}{\sqrt{2}} \, \tau_{3/2}(w)\,
\Big\{-(w-1)\epsilon_{\mu \nu \alpha \beta} \epsilon^{*\alpha}(v+ v^{\prime})^\beta+(\epsilon^* \cdot v)\epsilon_{\mu \nu \alpha \beta} v^\alpha v^{\prime \beta} \nn \\
&+&2 \epsilon_{\tau \sigma \alpha \beta}\epsilon^{*\sigma} v^\alpha v^{\prime \beta} \left[g^\tau_\mu v_\nu -g^\tau_\nu v_\mu \right] \nn \\
&+&i \left[(1+w)\left(\epsilon^*_\nu(v-v^\prime)_\mu-\epsilon^*_\mu(v-v^\prime)_\nu \right) -3 (\epsilon^* \cdot v)(v_\mu v^\prime_\nu-v_\nu v^\prime_\mu) \right] \Big\} \,\,\, ;  \label{T-D1} \\ \nn \\
<D_2^*(p^\prime,\,\epsilon)|{\bar c} \gamma_\mu(1-\gamma_5) b| B(p)>&=&\sqrt{m_B m_{D_2^*}} \, \sqrt{3} \,\tau_{3/2}(w) \, \Big\{-i \epsilon_{\mu \beta \tau \sigma}\left(\epsilon^{*\alpha \beta} v_\alpha \right) v^\tau v^{\prime \sigma} \nn \\ &+&
\left(\epsilon^{*\alpha \beta} v_\alpha \right) v_\beta v^\prime_\mu-(1+w) \left(\epsilon^{*\alpha}_\mu  v_\alpha \right)
 \Big\} \,\,\, ;  \label{V-A-D2star} \\
 <D_2^*(p^\prime,\,\epsilon)|{\bar c} \sigma_{\mu \nu}(1-\gamma_5) b| B(p)>&=&\sqrt{m_B m_{D_2^*}} \,\sqrt{3} \, \tau_{3/2}(w) \,
 \Big\{-\epsilon_{\mu \nu  \beta \tau} \left(\epsilon^{*\alpha \beta} v_\alpha \right) (v+ v^{\prime})^\tau \nn \\ &+& i
 \left(\epsilon^{*\alpha \tau} v_\alpha \right)\left[g^\tau_\mu (v+ v^{\prime})_\nu -g^\tau_\nu (v+ v^{\prime})_\mu \right] \Big\} \,\,\, .  \label{T-D2staqr}
 \eea
 In the previous formulae we have set $p=m_B \, v$, $p^\prime= m_{D^{**}}\, v^\prime$ and  $w=v \cdot v^\prime$;  $\epsilon$ is the polarization vector (tensor)  of the spin 1 (spin 2) $D^{**}$ meson.

The results for the SM, NP and interference contribution  to the differential distributions in (\ref{dgammadq2-generic})  are given below for each of the four excited mesons.
The relation between the squared momentum transfer $q^2$ and $w$ is  $q^2=m_B^2+m_{D^{**}}^2-2m_B m_{D^{**}}w$, with $m_{D^{**}}$ the mass of the charmed meson produced in the decay.
The lepton mass has been taken into account, hence the formulae also hold  for $\tau$.
\begin{itemize}
\item
$B \to D^*_0 \ell {\bar \nu}_\ell$:
\bea
{d \tilde \Gamma \over dq^2}(B \to D^*_0 \ell \bar \nu_\ell)\Big|_{SM} &=&
4 m_B m_{D_0^*} [\tau_{1/2}(w)]^2(w-1)\Big\{q^2 \left(1-\frac{m_\ell^2}{q^2} \right)+ \left(1+\frac{2m_\ell^2}{q^2} \right)\left[(m_B^2+m_{D_0^*}^2)w-2m_B m_{D_0^*}\right] \Big\}\nn\\
{d \tilde \Gamma \over dq^2}(B \to D^*_0 \ell \bar \nu_\ell)\Big|_{NP} &=&32 |\epsilon_T|^2
m_B m_{D_0^*} [\tau_{1/2}(w)]^2(w^2-1)\left(1+\frac{2m_\ell^2}{q^2} \right)(m_B^2+m_{D_0^*}^2-2m_B m_{D_0^*}w)
\label{Dstar0NP}
\\
{d \tilde \Gamma \over dq^2}(B \to D^*_0 \ell \bar \nu_\ell)\Big|_{INT} &=&-48  Re(\epsilon_T)
m_B m_{D_0^*} [\tau_{1/2}(w)]^2(w^2-1)m_\ell(m_B-m_{D_0^*}) \nn
\eea

\item
 $B \to D^\prime_1 \ell {\bar \nu}_\ell$:
\bea
{d \tilde \Gamma \over dq^2}(B \to D^\prime_1 \ell \bar \nu_\ell)\Big|_{SM} &=&4m_B m_{D_1^\prime} [\tau_{1/2}(w)]^2(w-1) \nn \\
&&\Big\{q^2 \left(1-\frac{m_\ell^2}{q^2} \right)(2w-1)+ \left(1+\frac{2m_\ell^2}{q^2} \right)\left[(m_B^2+m_{D_1^\prime}^2)3w-2m_B m_{D_1^\prime} (2w^2+1)\right] \Big\}\nn\\
{d \tilde \Gamma \over dq^2}(B \to D^\prime_1 \ell \bar \nu_\ell)\Big|_{NP} &=& 32 |\epsilon_T|^2
m_B m_{D_1^\prime} [\tau_{1/2}(w)]^2(w-1)\left(1+\frac{2m_\ell^2}{q^2} \right)
\nn \\
&&\Big\{(m_B^2+m_{D_1^\prime}^2)(5w-1)-2m_B m_{D_1^\prime}[4+w(w-1)] \Big\}
\label{D1primeNP}
\\
{d \tilde \Gamma \over dq^2}(B \to D^\prime_1 \ell \bar \nu_\ell)\Big|_{INT} &=& 48 Re(\epsilon_T)
m_B m_{D_1^\prime} [\tau_{1/2}(w)]^2(w-1)m_\ell[m_B(w-5)+m_{D_1^\prime}(5w-1)] \nn
\eea

\item
$B \to D_1 \ell {\bar \nu}_\ell$:
\bea
{d \tilde \Gamma \over dq^2}(B \to D_1 \ell \bar \nu_\ell)\Big|_{SM} &=& m_B m_{D_1} [\tau_{3/2}(w)]^2(w-1)(1+w)^2 \nn \\
&&\Big\{ q^2\left(1-\frac{m_\ell^2}{q^2} \right)(w-2)+\left(1+\frac{2m_\ell^2}{q^2} \right)\left[(m_B^2+m_{D_1}^2)3w-2m_B m_{D_1} (w^2+2)\right] \Big\} \nn
\\
{d \tilde \Gamma \over dq^2}(B \to D_1 \ell \bar \nu_\ell)\Big|_{NP} &=&16 |\epsilon_T|^2 m_B m_{D_1} [\tau_{3/2}(w)]^2(w-1)(1+w)^2 \left(1+\frac{2m_\ell^2}{q^2} \right) \nn \\
&&
\Big\{\left[(m_B^2+m_{D_1}^2)(2w-1)-2m_B m_{D_1} (w^2-w+1)\right] \Big\}
\label{D1NP}\\
{d \tilde \Gamma \over dq^2}(B \to D_1 \ell \bar \nu_\ell)\Big|_{INT} &=& 24 Re(\epsilon_T) m_B m_{D_1} [\tau_{3/2}(w)]^2(w-1)(1+w)^2 m_\ell [m_B(w-2)+m_{D_1}(2w-1)] \nn
\eea
\item
 $B \to D^*_2 \ell {\bar \nu}_\ell$:
\bea
{d \tilde \Gamma \over dq^2}(B \to D^*_2 \ell \bar \nu_\ell)\Big|_{SM} &=& m_B m_{D^*_2} [\tau_{3/2}(w)]^2(w-1)(1+w)^2 \nn  \\
&&\Big\{ q^2\left(1-\frac{m_\ell^2}{q^2} \right)(3w+2)+\left(1+\frac{2m_\ell^2}{q^2} \right)\left[(m_B^2+m_{D^*_2}^2)5w-2m_B m_{D^*_2} (3w^2+2)\right] \Big\}
\nn \\
{d \tilde \Gamma \over dq^2}(B \to D^*_2 \ell \bar \nu_\ell)\Big|_{NP} &=&16 |\epsilon_T|^2 m_B m_{D^*_2} [\tau_{3/2}(w)]^2(w-1)(1+w)^2\left(1+\frac{2m_\ell^2}{q^2} \right)  \nn \\
&&\Big\{\left[(m_B^2+m_{D^*_2}^2)(1+4w)-2m_B m_{D^*_2} (3+w+w^2)\right] \Big\}
\label{D2starNP}
\\
{d \tilde \Gamma \over dq^2}(B \to D^*_2 \ell \bar \nu_\ell)\Big|_{INT} &=&-24 Re(\epsilon_T) m_B m_{D^*_2} [\tau_{3/2}(w)]^2(w-1)(1+w)^2m_\ell[m_B(4+w)- m_{D^*_2}(1+4w) ]
\nn
\eea
\end{itemize}
The differential decay rates are obtained multiplying the above functions by the coefficient $C(q^2)$ in (\ref{C-factor}).

\end{document}